\documentclass[aps,prd,nofootinbib]{revtex4}
\usepackage[colorlinks=true, pdfstartview=FitV, linkcolor=blue, citecolor=red, urlcolor=magenta]{hyperref}
\usepackage{graphicx}
\usepackage{latexsym}
\usepackage{amsmath}
\usepackage{amsfonts}
\usepackage{amssymb}
\usepackage{verbatim}
\usepackage{braket}
\usepackage{extarrows}
\usepackage{hyperref}

\newcommand{\be}{\begin{equation}}
\newcommand{\ee}{\end{equation}}
\newcommand{\bea}{\begin{eqnarray}}
\newcommand{\eea}{\end{eqnarray}}

\newcommand{\ben}{\begin{eqnarray}}
\newcommand{\een}{\end{eqnarray}}

\begin{document}

\title{Loop correction to the scalar Casimir energy density and generation of topological mass due to a helix boundary condition in a scenario with Lorentz violation}

\author{$^{1}$A. J. D. Farias Junior}
\email{antonio.farias@academico.ufpb.br}

\author{$^{1}$Herondy F. Santana Mota}
\email{hmota@fisica.ufpb.br}

\affiliation{$^{1}$Departamento de F\' isica, Universidade Federal da Para\' iba,\\  Caixa Postal 5008, Jo\~ ao Pessoa, Para\' iba, Brazil.}


\begin{abstract}
In this paper the effective potential approach in quantum field theory is used in order
to investigate self-interaction loop correction to the Casimir energy density and generation of topological mass
for both massless and massive real scalar fields. It is assumed that the scalar field obeys a helix boundary condition. 
In addition, it is also considered a CPT-even aether-type violation of the Lorentz symmetry. In the absence of the Lorentz violation
we obtain analytical expressions for the loop correction to the Casimir energy density and to the mass of the scalar field. The same expressions are
also obtained assuming the Lorentz violation in each of the spacetime directions. We also show some graphs that exhibit how the loop correction and the Lorentz violation
affect the the Casimir energy density and the mass of the scalar field.

\end{abstract}
 \maketitle


\section{Introduction}
\label{intro}
%
One interesting and intriguing phenomenon which is of pure quantum nature is the
Casimir effect. It was first predicted by H. Casimir in (1948) \cite{Casimir:1948dh} who noticed that 
the effect consists in a force of attraction that arises between two neutral parallel
and perfectly conducting plates, placed in vacuum within a very short distance from each other.
This force of attraction is described in the framework of a quantum
electromagnetic field and it is the result of modifications on the vacuum fluctuations. Since its
prediction, the Casimir effect has been confirmed by several high accuracy
experiments \cite{Bressi:2002fr, Lamoreaux:1996wh, PhysRevLett.81.5475, mohideen1998precision, MOSTEPANENKO2000, PhysRevA.78.020101, PhysRevA.81.052115}. 
Currently, it is widely known that other quantum fields such as scalar
and fermion fields also produce Casimir-like effects when subjected to external influences, i.e.,
boundary conditions, temperature corrections, gauge fields and so on. For instance, the Casimir effect associated with a real scalar field subjected to a helix boundary condition with temperature corrections is considered in Ref. \cite{Aleixo:2021cfy}, 
subjected to Robin boundary
conditions in Ref. \cite{Romeo:2000wt}, and in Ref. \cite{Maluf:2019ujv}
the Casimir energy for a real scalar field and the Elko neutral spinor field
in a field theory at a Lifshitz fixed point is obtained. A review on the Casimir effect can be found in Ref. \cite{bordag2009advances} (see also \cite{Mostepanenko:1997sw, Milton:2001yy}). 

The most
standard approach to study the Casimir effect is assuming that the Lorentz
symmetry is preserved. However, all the attempts to build a high energy scale theory seems to fail in preserving the Lorentz symmetry, even locally in the case gravity is taken into consideration. In some effective Lorentz symmetry violation scenarios, the spacetime may become
anisotropic in some direction (including time), where the direction is usually determined
by a constant unit 4-vector. The Lorentz violation is a topic in which a great deal of attention has been given in the past years.
 In the context of string theory, for instance, the
Lorentz symmetry violation is studied in Ref. \cite{Kostelecky:1988zi} and in the low-energy scale
case in Refs. \cite{Anisimov:2001zc, Carlson:2001sw, Hewett:2000zp, Bertolami:2003nm, Kostelecky:2002ca, Anchordoqui:2003ij, Bertolami:1997iy, Alfaro:1999wd, Alfaro:2001rb}. As a consequence of such modifications in the
spacetime introduced by the Lorentz symmetry violation, the energy modes of
the quantum field change and so it modifies the Casimir energy density. In the context
of string theory and low-energy scale scenarios with Lorentz violation, the Casimir energy is
investigated in \cite{Obousy:2008xi, Obousy:2008ca} and in \cite{Martin-Ruiz:2016ijc, Martin-Ruiz:2016lyy, Escobar:2020pes}, respectively.

As we have mentioned, boundary conditions play a crucial role
in the investigation of the Casimir effect. A very interesting boundary condition that modifies the quantum vacuum fluctuations of a quantum field 
is the one describing a helix. The Casimir energy arising as a consequence of the imposition of a
helix boundary condition on a real scalar field has been considered in Refs. \cite{QuantumSpringFromCasimir, QuantumSpring, QuantumSpringD+1},
and a generalization including thermal effects in Ref. \cite{Aleixo:2021cfy}. In the
present paper we consider a self-interacting real scalar field subjected to a helix boundary
condition. The choice of such boundary condition is motivated by the large
number of structures in nature with a helix geometry, to name two of it we
have DNA and cell membrane proteins. In addition, we consider a scenario
where we allow the Lorentz symmetry violation to take place. In order to investigate the
Casimir energy density in this context, we use the path integral formalism for the
quantized field and then construct the effective potential which can be
written in terms of a loop expansion. This formalism was developed by Jackiw
\cite{Jackiw:1974cv} and allows us to obtain the Casimir energy density and its self-interaction loop corrections, including loop corrections to the mass of the field, in which case a topological mass is generated if the field 
is subjected to some boundary condition. In this sense, in Ref. \cite{Toms:1979ij} it was considered a massless real scalar field with self-interaction and different boundary conditions to obtain loop correction to the Casimir energy density and to the mass. Moreover, we take similar steps as in
\cite{Cruz:2020zkc} where a massive real scalar field with a Lorentz violation was
considered. However, we first consider both massless and massive real scalar
fields subjected to a helix boundary condition preserving the Lorentz symmetry. Next, we assume that a CPT-even aether-type Lorentz symmetry violation takes place
and analyze how the system previously considered preserving the Lorentz symmetry changes. Note that the CPT-even aether-type Lorentz symmetry violation model has been considered in Refs. \cite{Carroll:2008pk, Chatrabhuti:2009ew, Gomes:2009ch, Cruz:2020zkc}.

This paper is organized as Follows: in Sec.\ref{sec2} we review the main aspects of the path integral formalism to obtain the effective potential in the case of a self-interacting real scalar field. In Sec.\ref{sec3}, in Minkowski spacetime, we consider a self-interacting scalar field subjected to a helix boundary condition and, using the Riemann zeta function technique, obtain the Casimir energy density, its loop correction and the topological mass. In Sec.\ref{sec4} we consider the same system but in a CPT-even aether-type Lorentz symmetry violation scenario. We investigate how the Casimir energy density, its loop correction and the topological mass are affected by the Lorentz violation in each of the spacetime directions. In Sec.\ref{sec5} we present our conclusions. Throughout this paper we use natural units in which both the Planck constant and
the speed of light are $\hslash=c=1$. 

%
\section{Self-interacting scalar field}
\label{sec2}
Let us in this section briefly review some of the main aspects of a self-interacting real scalar field model and the subsequent effective potential approach for loop corrections, which in our case will be taken up to second order. Thus, for a minimally coupled massive scalar field $\varphi$, in a four-dimensional Euclidian spacetime, the action for such a model including the self-interaction term, $\lambda\varphi^4$, is written as \cite{Toms:1979ij}
\begin{equation}
S_{\text{E}}\left[  \varphi\right]  =\int d^{4}x\left[  -\frac{1}{2}(\partial^{\mu}%
\varphi)(\partial_{\mu}\varphi)-U\left(  \varphi\right)  \right]  ,
\label{a1}%
\end{equation}
where $U\left(  \varphi\right)$ is the classical potential. The latter, including the renormalization counterterms, is  given by
\begin{equation}
U\left(  \varphi\right)  =\frac{m^{2}}{2}\varphi^{2}+\frac{\lambda}{4!}%
\varphi^{4}+\frac{C_{1}}{4!}\varphi^{4}+\frac{C_{2}}{2}\varphi^{2}+C_{3},
\label{a1.1}%
\end{equation}
with $m$ being the mass of the scalar field, $\lambda$ is a coupling constant
and $C_{i}$ are the constants which will take care of the renormalization of
the effective potential. They shall be determined latter for the physical system we are interested in.

The path integral approach which leads to the construction of the effective
potential is described in detail in
Ref. \cite{Toms:1979ij, Ryder:1985wq, greiner2013field} (see also \cite{Cruz:2020zkc, Porfirio:2019gdy}). Here we present only the main expressions necessary for our purposes. Thus, the effective
potential, which we denote by $V_{\text{eff}}\left(  \Phi\right)  $, is obtained by
allowing the field $\varphi$ to fluctuate about a fixed background field
$\Phi$, that is, $\varphi =  \Phi + \phi $, with $\phi$ representing quantum fluctuations. Up to second order, the effective potential is written as%
\begin{equation}
V_{\text{eff}}\left(  \Phi\right)  =V^{\left(  0\right)  }\left(  \Phi\right)
+V^{\left(  1\right)  }\left(  \Phi\right)  +V^{\left(  2\right)  }\left(
\Phi\right)  . \label{a1.2}%
\end{equation}
The zero order term $V^{\left(  0\right)  }\left(  \Phi\right) = U\left(  \Phi\right) $ is just the
classical potential, that is, the tree-level contribution to the effective
potential.
The next term, $V^{\left(  1\right)  }\left(  \Phi\right)  $, is the one-loop
correction to the classical potential and can be written in terms of a path integral as \cite{Toms:1979ij} 
\begin{align}
V^{\left(  1\right)  }\left(  \Phi\right)  =-\frac{1}{\Omega_{4}}%
\ln\int\mathcal{D}\varphi\exp\left\{  -\frac{1}{2}\int d^{4}x\varphi\left(
x\right)  \hat{A}\varphi\left(  x\right)  \right\}  ,\label{v1.1}
\end{align}
where the self-adjoint elliptic differential operator $\hat{A}$ is defined as
\begin{align}
 \hat{A}=U^{\prime\prime}\left(  \Phi\right)  -\square,
 \label{l}
\end{align}
being $\Omega_{4}$ the four-dimensional volume of the Euclidian spacetime, $U^{\prime
\prime}\left(  \Phi\right)  $ is the second derivative of the classical potential with
respect to the field, evaluated at the fixed background field $\Phi$, and
$\square$ is the D'Alembertian operador in Euclidean coordinates, that is,
\begin{equation}
\square=\partial_{\tau}^{2}+\mathbf{\nabla}^{2}. 
\label{doperator}%
\end{equation}
One should note that the Euclidian formalism is realized once we make a wick rotation in the real time coordinate $t$, introducing an imaginary time coordinate $\tau$, that is, $t=-i\tau$ \cite{Hawking1977, Aleixo:2021cfy}. The differential operator $\hat{A}$, after allowing the field $\varphi$ to fluctuate about $\Phi$, will act on the fluctuating part $\phi$.

It can be shown that the one-loop correction (\ref{v1.1}) is written in terms
of the generalized zeta function $\zeta\left(  s\right)$ as \cite{Toms:1979ij, Hawking1977}
\begin{equation}
V^{\left(  1\right)  }\left(  \Phi\right)  =-\frac{1}{2\Omega_{4}%
}\left[  \zeta^{\prime}\left(  0\right)  +\zeta\left(  0\right)  \ln\mu
^{2}\right]  , \label{z1}%
\end{equation}
where $\zeta\left(  0\right)  $ and $\zeta^{\prime}\left(  0\right)  $ are the
zeta function and its derivative with respect to $s$, evaluated at $s=0$, respectively. Note that $\mu$ is a constant parameter which will be removed by renormalization. The
eigenvalues of the operator $\hat{A}$ defined in Eq.~(\ref{l}), which we
denote by $a_{\sigma}$, are used in the construction of the generalized zeta function, that is,
\begin{equation}
\zeta\left(  s\right)  =\sum_{\sigma}a_{\sigma}^{-s}, \label{z2}%
\end{equation}
with $\sigma$ representing the set of quantum numbers associated with the eigenfunctions of the operator $\hat{A}$, and the sum sign stands for a summation or integration, depending on whether the quantum numbers are discrete or continuum, respectively. For practical
reasons, the two-loop contribution for the effective potential is calculated
from the two-loop graph and we can also write it in terms of the generalized zeta function if we are interested in calculating this contribution at $\Phi=0$, which is in fact the vacuum contribution \cite{Cruz:2020zkc, Porfirio:2019gdy}. We postpone the explicit form 
for the two-loop contribution until we investigate it.

The renormalization condition, which allows us to remove the term
$\zeta\left(  0\right)  \ln\mu^{2}$ in Eq.~(\ref{z1}) and in addition fix the
constant $C_{1}$ in Eq.~(\ref{a1.1}), is written in analogy to
Coleman-Weinberg, also fixing the coupling constant $\lambda$ \cite{Coleman:1973jx}, that is,
\begin{equation}
\left.  \frac{d^{4}V_{\text{eff}}}{d\Phi^{4}}\right\vert _{\Phi=0}=\lambda.
\label{2.8}%
\end{equation}
Note that in the case the field is massless, the constant $\lambda$ should be fixed at some energy scale $\Phi=M$ \cite{Toms:1979ij, Porfirio:2019gdy}.

The condition which yields the topological mass and fix the constant $C_{2}$ in
(\ref{a1.1}), is written as%
\begin{equation}
\left.  \frac{d^{2}V_{\text{eff}}}{d\Phi^{2}}\right\vert _{\Phi=0}=m^{2}, \label{2.9}%
\end{equation}
which we expect to be positive since it is connected with the mass of the scalar field. Eq. (\ref{2.9}) also provides the topological mass when we use the
renormalized effective potential instead of $V_{\text{eff}}$. One should point out that $\Phi=0$ in Eq. \eqref{2.9} is the value that minimizes the effective potential and represents its minimum as long as it obeys the extremum condition 
\begin{equation}
\left.  \frac{dV_{\text{eff}}}{d\Phi}\right\vert _{\Phi=0}=0. \label{ext}%
\end{equation}
The minimum of the effective potential at $\Phi=0$ is exactly the vacuum state we are interested in, to calculate the vacuum energy, its loop correction and the generation of topological mass.

Finally, to find the constant
$C_{3}$ one has to use an additional renormalization condition \cite{Cruz:2020zkc},
that is,%
\begin{equation}
\left.  V_{\text{eff}}\right\vert _{\Phi=0}=0. \label{2.11}%
\end{equation}
Note that the renormalization conditions \eqref{2.8},  \eqref{2.9} and  \eqref{2.11} are to be taken in the limit of Minkowski
spacetime.

With the necessary tools described above, we can proceed to
the investigation of the loop corrections to the classical potential and formation of topological mass when the fluctuating scalar field, $\phi$, is subjected to a helix boundary condition. In the next section, we shall first consider the fluctuating scalar field part, that is, $\phi$, subjected to a helix boundary condition in Minkowski spacetime. In Sec.\ref{sec4}, subjecting the field $\phi$ to the same boundary condition, we consider a scenario where there is a CPT-even aether-like Lorentz violation of symmetry \cite{Carroll:2008pk, Chatrabhuti:2009ew, Gomes:2009ch} and, consequently discuss the difference from the case without violation.

\section{Helix boundary condition in Minkowski spacetime}
\label{sec3}
As already mentioned, once the field $\varphi$ is expanded about the fix background $\Phi$, its fluctuating part $\phi$ will represent quantum oscillations. Thus, it is exactly the latter that will be sensible to any boundary condition imposed on it. In this sense, our interest in this section is to impose on the field $\phi$ a helix boundary condition and investigate, up to second order, loop corrections to the effective potential. As we will see below, the Casimir energy density as a consequence of the imposition of a helix boundary condition on the quantum scalar field $\phi$, at order $\lambda^0$, is provided by the first loop correction to the effective potential. We, then, verify that the results for the massive and massless scalar field cases are in agreement with the ones found in Refs. \cite{QuantumSpringFromCasimir, QuantumSpring, QuantumSpringD+1, Aleixo:2021cfy}, using other methods. Additionally, we also calculate the second loop correction to the effective potential, which translates in being the correction to the Casimir energy density at order $\lambda$ due to the self-interaction considered in Eq. \eqref{a1}. The generation of a correction, of order $\lambda$, to the mass $m$ of the field is also obtained. 

The set of eigenfunctions of the eigenvalue equation $\hat{A}\phi_{\sigma} = a_{\sigma}\phi_{\sigma}$ under the imposition of the helix boundary condition
\begin{equation}
\phi\left(  \tau,x+a,y,z\right)  = \phi\left(  \tau,x,y+h,z\right) ,
\label{he1}%
\end{equation}
is given by 
\begin{eqnarray}
\phi_{\sigma} = Ne^{-ik_{\tau}\tau + ik_xx + ik_yy + ik_zz},
\label{eifun}
\end{eqnarray}
%
where $N$ is a normalization constant\footnote{The normalization constant $N$ is irrelevant for our purposes, but its calculation can be found in detail in Ref. \cite{Aleixo:2021cfy}.} and $k_xa-k_yh = 2\pi n$, with $n=0, \pm 1, \pm 2, \pm 3, ...$\, \cite{QuantumSpringFromCasimir, QuantumSpring, QuantumSpringD+1, Aleixo:2021cfy}. Note that the parameter $h$ stands for the pitch of the helix while $a$ for its radius. The eigenvalues, $a_{\sigma}$, of the operator $\hat{A}$, are given by
\begin{eqnarray}
a_{\sigma}=k_{\tau}^{2} + k_{z}^{2}  + k_{x}^{2}+\left(  \frac{k_{x}a}{h}-\frac{2\pi n}{h}\right)^{2} +M_{\Phi}^{2}%
,\qquad\qquad n=0,\pm1,\pm2,...,\label{aeigen} \label{aeigen2}%
\end{eqnarray}
where $M_{\Phi}^{2}=m^{2}+\frac{\lambda}{2}\Phi^{2}$ and $\sigma = (k_{\tau}, k_x, k_z, n)$ stands for the set of quantum numbers.
In order to construct the generalized zeta function from the eigenvalues
(\ref{aeigen}) and obtain a practical form for it, we follow the steps
presented in \cite{Cruz:2020zkc, Porfirio:2019gdy}.


\subsubsection{Effective potential up to one-loop correction}
%
The standard procedure to build the generalized zeta function \eqref{z2} for our case is to use the eigenvalues obtained in Eq. (\ref{aeigen}). Thereby, we have
\begin{equation}
\zeta\left(  s\right)  =\frac{\Omega_{3}}{\left(  2\pi\right)  ^{3}}\int
dk_{\tau}\ dk_{x}\ dk_z\sum_{n=-\infty}^{+\infty}\left[  k_{\tau}^{2} + k_{z}^{2}  + k_{x}^{2}+\left(  \frac{k_{x}a}{h}-\frac{2\pi n}{h}\right)^{2}  +M_{\Phi}^{2}\right]  ^{-s}, \label{z3}%
\end{equation}
where $\Omega_{3}$ is a 3-dimensional volume associated with the coordinates
$\tau,x,z$, necessary to make the integrals dimensionless. By using the identity,%
\begin{equation}
w^{-s}=\frac{2}{\Gamma\left(  s\right)  }\int_{0}^{\infty}d\tau\ \tau
^{2s-1}e^{-w\tau^{2}}, \label{i1}%
\end{equation}
we are able to perform the resulting Gaussian integrals in $k_{\tau}$, $k_{x}$ and $k_{z}$. In this case, we are left with
\begin{equation}
\zeta\left(  s\right)  =\frac{\Omega_{4}}{\left(  2\pi\right)  ^{2}d}%
\frac{\pi^{\frac{1}{2}}}{\Gamma\left(  s\right)  }\sum_{n=-\infty}^{+\infty}\int
_{0}^{\infty}d\tau\ \tau^{2s-4}\exp\left\{  -\tau^{2}\left[  M_{\Phi}%
^{2}+\frac{4\pi^2}{d^{2}}n^{2}\right]  \right\}  ,
\label{z}%
\end{equation}
where we have defined $\Omega_{4}=\Omega_{3}h$ and %
\begin{equation}
d^{2}=a^{2}+h^{2}. \label{d1}%
\end{equation}
From the well known integral representation of the gamma function $\Gamma\left(
z\right)  $ \cite{Abramowitz},%
\[
\Gamma\left(  z\right)  =2\int_{0}^{\infty}d\mu\ \mu^{2z-1}e^{-\mu^{2}},
\]
the zeta function (\ref{z}) can be written in terms only of the summation in $n$ as%
\begin{align}
\zeta\left(  s\right)  =\frac{\Omega_{4}\pi^{\frac{1}{2}}}{2\left(  2\pi\right)
^{2s-1}d^{4-2s}}\frac{\Gamma\left(  s-\frac{3}{2}\right)  }{\Gamma\left(  s\right)
}\sum_{n=-\infty}^{+\infty}\left[ \left(\frac{M_{\Phi} d}{2\pi}\right)^2 + n^{2} \right]  ^{\frac{3}{2}%
-s}.
\label{z2222}
\end{align}
The sum in $n$ above can be performed by making use of the following analytic continuation of the inhomogeneous, generalized Epstein function \cite{Elizalde:1995hck, Feng:2013zza}:
\begin{eqnarray}
 \sum_{n=-\infty}^{+\infty}\left[  \left(  n+\beta\right)  ^{2}+\nu
^{2}\right]  ^{-s}=\frac{\pi^{\frac{1}{2}}\nu^{1-2s}}{\Gamma\left(  s\right)
}\left\{  \Gamma\left(  s-\frac{1}{2}\right) + 4(\pi\nu)^{s-\frac{1}{2}}\sum_{j=1}j^{s-\frac{1}{2}%
}\cos\left(  2\pi j\beta\right)  K_{ \frac{1}{2}-s  }\left(
2\pi j\nu\right)  \right\}  ,
\label{summ}
\end{eqnarray}
considering $\beta=0$. Note that $K_{\mu}(x)$ is the modified Bessel function of the second kind or, as it is also known, the Macdonald function \cite{Abramowitz}. Thus, Eq. \eqref{z2222} is now written as
\begin{eqnarray}
\zeta(s) = \frac{\Omega_{4}M_{\Phi}^{4-2s}}{16  \pi^{2}}\frac{\Gamma\left(  s-2\right)  }{\Gamma\left(  s\right)} + \frac{\Omega_4}{2^{s}\pi^2\Gamma(s)}\left(\frac{M_{\Phi}}{d
}\right)^{2-s}\sum_{j=1}^{\infty}j^{s-2}K_{\left(  2-s\right)  }\left(  jdM_{\Phi}\right).
\label{gzeta}
\end{eqnarray}
With the generalized zeta function expression above we are now able to calcule the one-loop correction \eqref{z1} to the effective potential. In order to do that, one needs to take Eq. \eqref{gzeta} and its derivative with respect to $s$, at $s=0$. The evaluation of the generalized zeta function \eqref{gzeta}, at $s=0$, provides a finite result that comes from the first term on the r.h.s. The second term goes to zero since $\Gamma(s=0)\rightarrow\infty$. The result is, then, written as
%
\begin{eqnarray}
\zeta(0) = \frac{\Omega_{4}M_{\Phi}^{4}}{32  \pi^{2}},
\label{gzetazero}
\end{eqnarray}
which is consistent with previous works \cite{Toms:1979ij, Cruz:2020zkc, Porfirio:2019gdy}. The above expression for $\zeta(0)$, as we shall see, is to be removed later by the renormalization process, when we calculate $C_1$ through the condition \eqref{2.8}.  Eq. \eqref{gzetazero} gives a boundary independent contribution to the effective potential \eqref{a1.2}, when we take $d\rightarrow\infty$. 

The derivative of the zeta function (\ref{gzeta}) with respect to $s$, taken at $s=0$, is found to be
\begin{eqnarray}
\zeta'(0) =- \frac{\Omega_{4}M_{\Phi}^{4}}{32  \pi^{2}}\left[\ln(M_{\Phi}^2) - \frac{3}{2}\right] + \frac{\Omega_{4}M_{\Phi}^{4}}{\pi^{2}}\sum_{j=1}^{\infty}f_2\left(  jdM_{\Phi}\right),
\label{gzetaone}
\end{eqnarray}
where we have defined the function
\begin{equation}
f_{\gamma}\left(  x\right)  = \frac{K_{\gamma}\left(
x\right)  }{x^{\gamma}}. \label{f}%
\end{equation}
The result obtained in Eq. \eqref{gzetaone} is the one that contributes with nontrivial effects to the renormalized effective potential, due to the boundary condition \eqref{he1}.

Collecting the expressions \eqref{gzetazero} and \eqref{gzetaone} into Eq. \eqref{z1}, we found that the one-loop correction to the effective potential is written as
\begin{equation}
V^{\left(  1\right)  }\left(  \Phi\right)  =\frac{M_{\Phi}^{4}}%
{64\pi^{2}}\left[  \ln\left(  \frac{M_{\Phi}^{2}}{\mu^{2}}\right)  -\frac
{3}{2}\right]  - \frac{M_{\Phi}^{4}}{2\pi^{2}}\sum_{j=1}^{\infty}f_{2}\left( jM_{\Phi
}\sqrt{a^2 + h^2}\right)  . \label{v1m}%
\end{equation}
Consequently, the nonrenormalized effective potential up to one-loop correction is obtained from \eqref{a1.2} as
\begin{align}
  V_{\text{eff}}\left(  \Phi\right)  & =\frac{\lambda}{4!}\Phi^{4}+\frac{C_{1}}%
{4!}\Phi^{4}+\frac{m^{2}}{2!}\Phi^{2}+\frac{C_{2}}{2!}\Phi^{2}+C_{3}\nonumber\\
&  +\frac{M_{\Phi}^{4}}{64\pi^{2}}\left[  \ln\left(  \frac{M_{\Phi}^{2}}%
{\mu^{2}}\right)  -\frac{3}{2}\right]  - \frac{M_{\Phi}^{4}}{2\pi^{2}}\sum_{j=1}^{\infty}f_{2}\left( jM_{\Phi
}\sqrt{a^2 + h^2}\right)   . \label{v1un}%
\end{align}
It remains to be found now the renormalization constants $C_1$, $C_2$ and  $C_3$. Thereby, in order to calculate $C_1$, for instance, we make use of the renormalization condition in Eq. (\ref{2.8}), taking the Minkowski limit $d\rightarrow\infty$ afterwards \cite{Toms:1979ij, Cruz:2020zkc, Porfirio:2019gdy}. This gives
\begin{equation}
\frac{C_{1}}{4!}=\frac{\lambda^{2}}{256\pi^{2}}\ln\left(  \frac{\mu^{2}%
}{m^{2}}\right), \label{c1m}%
\end{equation}
which is the one responsible for subtracting the term containing $\ln(\mu^2)$ coming from the generalized zeta function \eqref{z1}. Similarly, the conditions (\ref{2.9}) and (\ref{2.11}) yields%
\begin{eqnarray}
\frac{C_{2} }{2!} =\frac{\lambda m^{2}}{64\pi^{2}}+\frac{\lambda m^{2}}{64\pi^{2}}%
\ln\left(  \frac{\mu^{2}}{m^{2}}\right),\label{c2m}
\label{c3m}
\end{eqnarray}
and
\begin{eqnarray}
C_{3} =\frac{3m^{4}}{128\pi^{2}}+\frac{m^{4}}{64\pi^{2}}\ln\left(
\frac{\mu^{2}}{m^{2}}\right), \label{c3m}%
\end{eqnarray}
respectively. The substitution of the renormalization constants just found above in the effective potential \eqref{v1un} provides the renormalized effective potential at one-loop level, i.e.,
\begin{align}
  V_{\text{eff}}^{R}\left(  \Phi\right) & =\frac{\lambda}{4!}\Phi^{4}+\frac{m^{2}}%
{2!}\Phi^{2}+\frac{\lambda m^{2}\Phi^{2}}{64\pi^{2}}\left[  \ln\left(
\frac{M_{\Phi}^{2}}{m^{2}}\right)  -\frac{1}{2}\right]  +\frac{m^{4}}{64\pi^{2}}\ln\left(  \frac{M_{\Phi}^{2}}{m^{2}}\right)\nonumber\\
&  +\frac{\lambda^{2}\Phi^{4}}{256\pi^{2}}\left[  \ln\left(  \frac{M_{\Phi}^{2}%
}{m^{2}}\right)  -\frac{3}{2}\right]  -\frac{M_{\Phi}^{4}}{2\pi^{2}}%
\sum_{j=1}^{\infty}f_{2}\left( jM_{\Phi
}\sqrt{a^2 + h^2}\right) . \label{vrm}%
\end{align}
The closed expression above for the renormalized effective potential at one-loop level correction clearly presents a dependence on the helix boundary condition parameters, $\sqrt{a^2 + h^2}$, which affects the last term on the r.h.s. This term goes to zero in the Minkowski limit, $d\rightarrow\infty$, remaining only the well known Coleman-Weinberg renormalized effecitve potential. This is easy to realize once we use the asymptotic limit for large arguments of the Macdonald function, namely, $K_{\mu}(w)\simeq\left(\frac{\pi}{2w}\right)^{\frac{1}{2}}e^{-w}$ \cite{Abramowitz}. 

We wish now to calculate the Casimir energy density from the renormalized effective potential \eqref{vrm}. This is straightforward once we know the vacuum state, which in our case is at $\Phi=0$, as anticipated below Eq. \eqref{ext}. Thus, by taking Eq. \eqref{vrm} at the vacuum state we have
\begin{equation}
\mathcal{E}_{\text{C}}=\left.  V_{\text{eff}}^{R}\left(  \Phi\right)  \right\vert _{\Phi=0}=-\frac{m^{4}}{2\pi^{2}}%
\sum_{j=1}^{\infty}f_{2}\left( jm\sqrt{a^2 + h^2}\right)  . \label{casm}%
\end{equation}
We, therefore, conclude that by using the effective potential approach the phenomenon of the Casimir effect arises right from the first quantum correction to the effective potential, which is of order $\lambda^0$, as it should be.

The Casimir energy density associated with a massless scalar field, in our case, is obtained from (\ref{casm}) by taking the limit for small arguments of the Macdonald function, that is, $K_{\mu}\left(  x\right)  \approx\frac{\Gamma\left(\mu\right)  }{2}\left(  \frac{2}{x}\right)  ^{\mu}$ \cite{Abramowitz}. This gives
\begin{equation}
\mathcal{E}_{\text{C}}=-\frac
{\pi^{2}}{90(a^2 + h^2)^{2}}, \label{cas0}%
\end{equation}
where we have used the standard Riemann zeta function $\zeta\left(  4\right)  = \frac{\pi^{4}}{90}$,
\cite{Elizalde:1995hck, Elizalde1994book}. The results in Eqs. \eqref{casm} and \eqref{cas0}  for the Casimir energy densities are in agreement with the ones obtained in Refs. \cite{QuantumSpringFromCasimir, QuantumSpring, QuantumSpringD+1, Aleixo:2021cfy} by different methods. 

\begin{figure}[h!]
	\centering
	{\includegraphics[width=8cm]{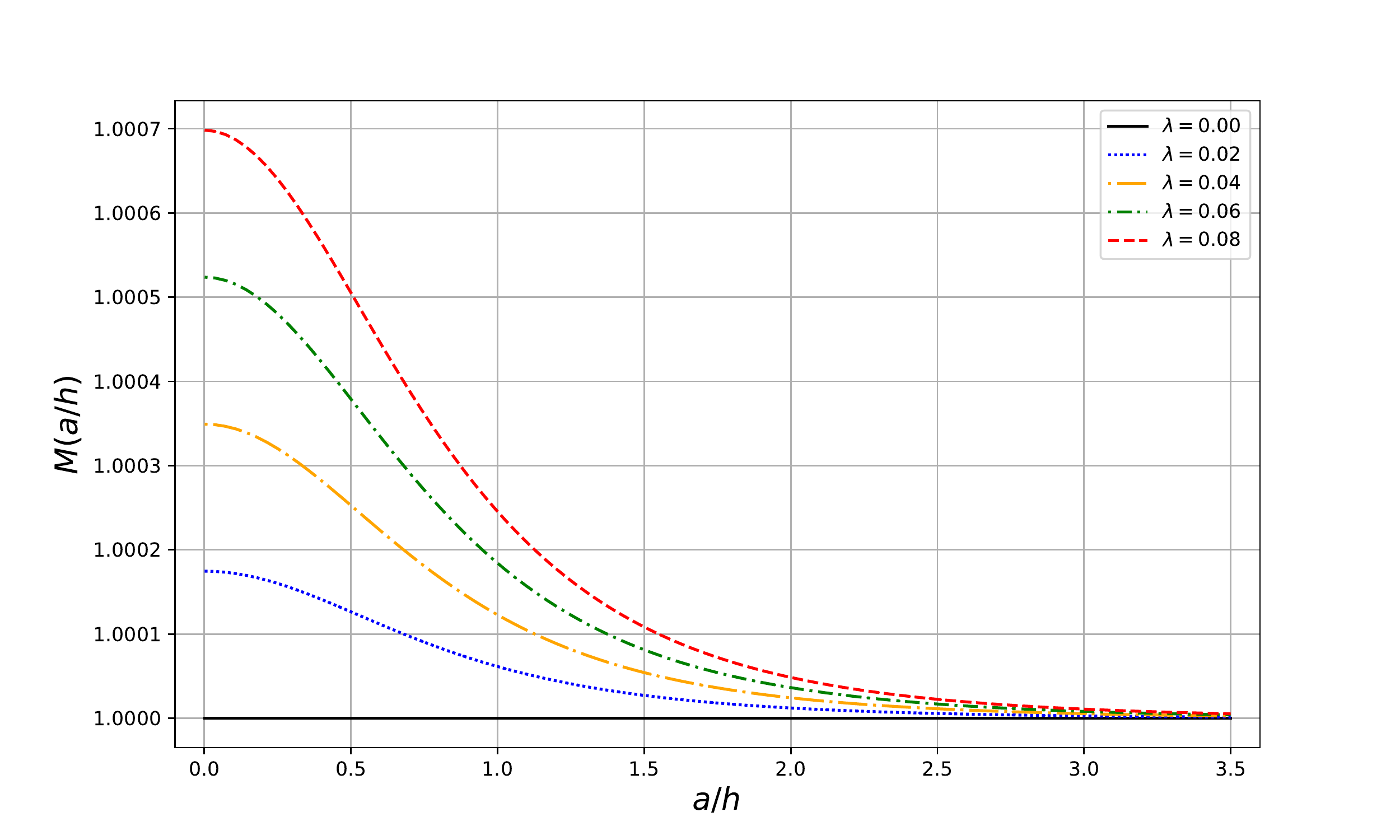} }%
	\qquad
	{\includegraphics[width=8cm]{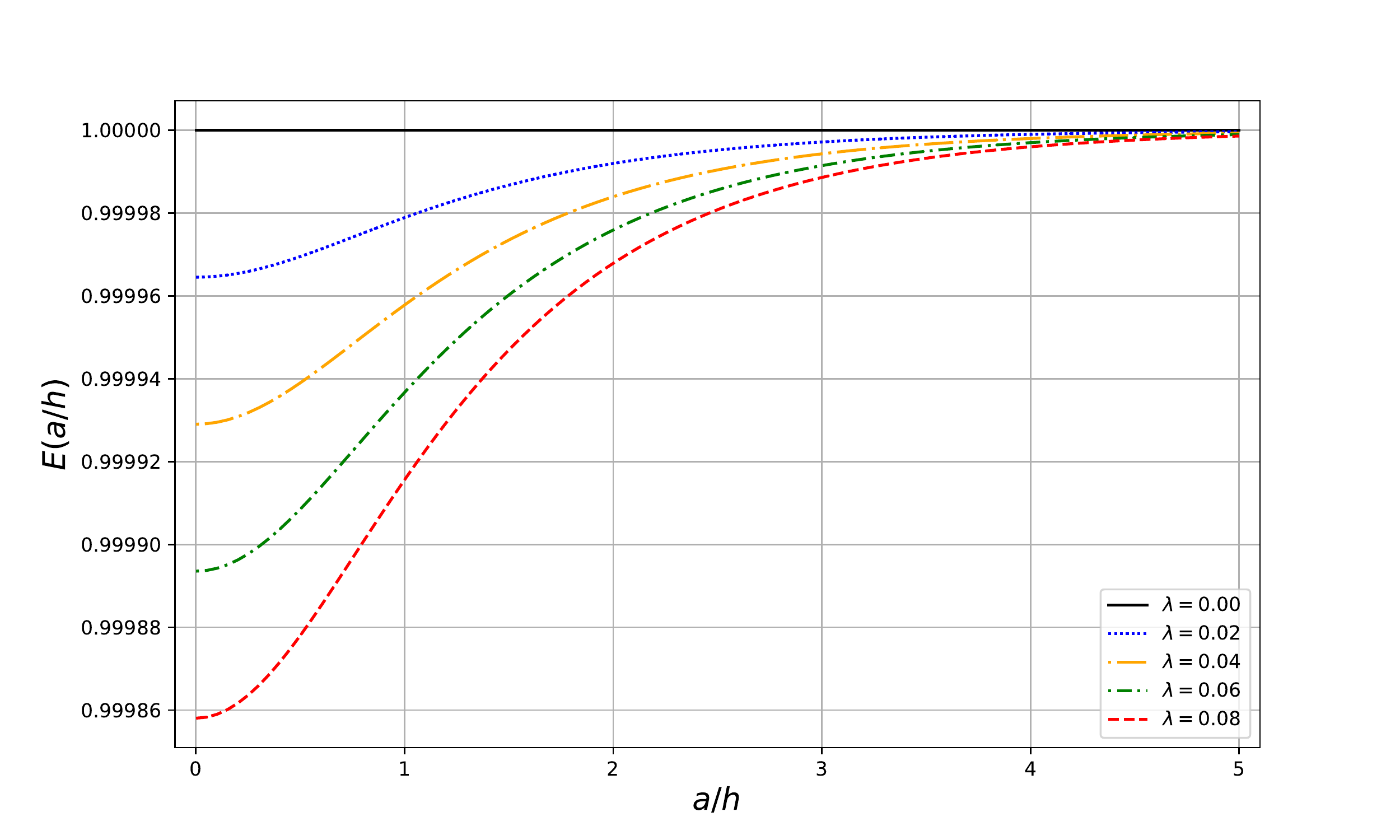} }%
	\caption{For different values of $\lambda$, the plot on the left shows the curves of $M=\frac{m_{\text{T}}}{m}$ with respect to $\frac{a}{h}$ from Eq. \eqref{mtm}, while the plot on the right shows the curves of $E=\frac{\mathcal{E}_{C}^{\lambda}}{\mathcal{E}_C}$ with respect to $\frac{a}{h}$ from Eq. \eqref{caslc}. For both plots it is assumed $mh=1$.}
	\label{figure1}%
\end{figure}

We are in fact interested in calculating the loop correction to the Casimir energy densities \eqref{casm} and \eqref{cas0}. This is achieved by obtaining the two-loop correction to the effective potential \eqref{v1un} which is nonzero since the influence of a helix boundary condition is present. Before doing that, let us investigate the influence, at one-loop level, of the boundary condition in the mass, $m$, of the self-interacting scalar field. Thereby, the condition (\ref{2.9}), along with the renormalized effective potential \eqref{vrm}, allow us to obtain 
\begin{equation}
m_{\text{T}}^{2}=m^{2}\left[  1+\frac{\lambda}{4\pi^{2}}\sum_{j=1}^{\infty}f_{1}\left( jm\sqrt{a^2 + h^2}\right)   \right]  . \label{mtm}%
\end{equation}
It is evident that the mass, $m$, of the field has gained a correction of order $\lambda$ that is dependent of the boundary condition codified by the radius, $a$, and pitch, $h$, of the helix. This correction arises at one-loop level and, as a consequence, the whole expression in Eq. \eqref{mtm} is called topological mass. On the left of Fig.\ref{figure1}, Eq. \eqref{mtm} is plotted as a dimensionless mass $\frac{m_{\text{T}}}{m}$ with respect to the dimensionless length $\frac{a}{h}$, for different values of the coupling constant $\lambda$ and assuming $mh =1$. We clearly see that the effect of the one-loop correction is to increase the mass $m$. However, this correction goes to zero at large values of $\frac{a}{h}$ making the mass $m$ dominant. 
%

In the Minkowski limit, $d\rightarrow\infty$, the expression \eqref{mtm} reduces solely to the mass, $m$, of the field while in the case of  a self-interacting massless scalar field it reduces to 
\begin{equation}
m_{\text{T}} ^{2}=\frac{\lambda}{24(a^2 + h^2)}, \label{m0tm}%
\end{equation}
where we have used $\zeta\left(  2\right)  =\frac{\pi^{2}}{6}$ \cite{Elizalde:1995hck, Elizalde1994book}. The expression in Eq. \eqref{m0tm} is purely topological mass and means that, even though the field starts classically with no mass, at one-loop level quantum correction, as a consequence of the boundary condition \eqref{he1}, it gains a mass-like contribution. 


%
\subsubsection{Two-loop correction}
It can be shown that the two-loop contribution to the effective potential
(\ref{a1.2}) is obtained by considering the two-loop irreducible graph
for one particle \cite{Jackiw:1974cv, Toms:1979ij}. As we are interested in calculating the two-loop contribution at $\Phi=0$, the only nonzero contribution\footnote{In Ref. \cite{ Toms:1979ij} it has been shown that from the two possible graphs to this loop correction, only one gives a nonzero contribution at $\Phi=0$. } coming from the graphs can be written in terms of the generalized zeta function \eqref{gzeta} as \cite{Cruz:2020zkc, Porfirio:2019gdy}
\begin{equation}
\left.  V^{\left(  2\right)  }\left(  \Phi\right)  \right\vert _{\Phi
=0}=\lim_{s\rightarrow1}\frac{\lambda}{8}\left.  \left(  \frac{\zeta
_{R}\left(  s\right)  }{\Omega_{4}}\right)  ^{2}\right\vert _{\Phi=0},
\label{v2.2}%
\end{equation}
where $\zeta_{R}\left(  s\right)$ is defined as \eqref{gzeta} minus the first term on the r.h.s., that is,
\begin{equation}
\zeta_{R}\left(  s\right)  = \zeta\left(  s\right)  -  \frac{\Omega_{4}M_{\Phi}^{4-2s}}{16  \pi^{2}}\frac{\Gamma\left(  s-2\right)  }{\Gamma\left(  s\right)},
\label{zr1}%
\end{equation}
being the subtracted term divergent at $s=1$. Note that the second term on the r.h.s of \eqref{zr1} gives a contribution that is independent of the boundary condition parameters $a$ and $h$ and, as usual, should be dropped since it is divergent at $s=1$ \cite{Cruz:2020zkc, Porfirio:2019gdy}. Thus, taking into consideration \eqref{zr1} in \eqref{v2.2} we find
\begin{equation}
\left.  V^{\left(  2\right)  }\left(  \Phi\right)  \right\vert _{\Phi
=0}=\frac{\lambda m^{4}}{32\pi^{4}}\left[  \sum_{j=1}^{\infty}%
f_1\left(  jm\sqrt{a^2 + h^2}\right)  \right]  ^{2}, \label{v2.4}%
\end{equation}
which is a completely convergent loop correction to the Casimir energy density \eqref{casm} associated with a self-interacting massive scalar field. The massless scalar field case of Eq. \eqref{v2.4} is obtained by taking the limit of small argument for the Macdonald function. This gives
\begin{equation}
\left.  V^{\left(  2\right)  }\left(  \Phi\right)  \right\vert _{\Phi
=0}=\frac{\lambda}{1152(a^2 + h^2)^2}. \label{v2.4massless}%
\end{equation}
%

The Casimir energy densities \eqref{casm} and \eqref{cas0} with their loop correction above are, therefore, found to be
\begin{equation}
\mathcal{E}_{\text{C}}^{\lambda}= \mathcal{E}_{C}+ \frac{\lambda m^{4}}{32\pi^{4}}\left[  \sum_{j=1}^{\infty}%
f_1\left(  jm\sqrt{a^2 + h^2}\right)  \right]  ^{2} , 
\label{caslc}%
\end{equation}
for the massive case and 
\begin{equation}
\mathcal{E}_{\text{C}}^{\lambda}=\mathcal{E}_{C}+ \frac{\lambda}{1152(a^2 + h^2)^2}, 
\label{caslcmassless}%
\end{equation}
for the massless case, respectively. Note that the first term on the r.h.s of Eq. \eqref{caslc} is given by \eqref{casm}, while the first term on the r.h.s of Eq. \eqref{caslcmassless} is given by \eqref{cas0}. Note also that in both cases, in the absence of self-interaction, that is, $\lambda=0$, we recover the already known results in Eqs. \eqref{casm} and \eqref{cas0}. On the right of Fig.\ref{figure1}, Eq. \eqref{caslc} is plotted as a dimensionless energy density $\frac{\mathcal{E}_{C}^{\lambda}}{\mathcal{E}_C}$ with respect to the dimensionless length $\frac{a}{h}$, for different values of the coupling constant $\lambda$ and taking $mh =1$. We can see that the effect of the loop correction is to decrease the Casimir energy density. However, this correction goes to zero at large values of $\frac{a}{h}$ making the Casimir energy density \eqref{casm} dominant.

%
\section{Self-interacting scalar field with Lorentz symmetry violation}
\label{sec4}
In this section it is considered a massive self-interacting real scalar field with
a CPT-even aether-type Lorentz symmetry violation \cite{Carroll:2008pk, Chatrabhuti:2009ew, Gomes:2009ch}.
This system is described by the following Euclidean action:%
\begin{equation}
S_{\text{E}}\left[  \varphi\right]  =\int d^{4}x\left[  -\frac{1}{2}(\partial^{\mu}%
\varphi)(\partial_{\mu}\varphi)
-\frac{1}{2}\chi\left(u^{0}\partial_{0
}\varphi\right)^2 + \frac{1}{2}\chi \left(u^{i}\partial_{i}\varphi\right)^2 
 -U\left(
\varphi\right)  \right]  ,
\label{LVAC}
\end{equation}
where $U\left(  \varphi\right)  $ is given in Eq.~(\ref{a1.1}),
$u^{\mu}$ is a unit 4-vector in Euclidean spacetime that points in the direction along which the Lorentz symmetry violation is supposed to occur, and the dimensionless parameter $\chi$ is a small number that characterizes the violation. Note that the term which is responsible for breaking the Lorentz symmetry is given by $\mathcal{L}_{\text{LV}}=\frac{1}{2}\chi(u^{\mu}\partial_{\mu})^2$, in Minkowski spacetime. However, due to the Wick rotation $t=-i\tau$, the time component part of this term gains a minus sign that is present in the Euclidean action \eqref{LVAC}.

The action \eqref{LVAC}, defining the self-interacting scalar field theory with Lorentz violation, allows us to re-write the operator $\hat{A}$, originally presented in
Eq.~(\ref{l}), in the form \cite{Cruz:2020zkc}%
\begin{equation}
\hat{A}=U''(\Phi) - \chi u^{0}u^{0}\partial_{0}\partial_{0 }+ \chi u^{i}u^{i}\partial_{i}\partial_{i}-\square. \label{l3}%
\end{equation}
Note that, in the absence of Lorentz violation, i.e., $\chi=0$, the operator above reduces to the form in Eq. \eqref{l}. The set of eigenfunctions of the eigenvalue equation $\hat{A}\phi_{\sigma} = a_{\sigma}\phi_{\sigma}$ under the imposition of the helix boundary condition \eqref{he1} is still the one in Eq. \eqref{eifun}. However, the set of eigenvalues of the operator $\hat{A}$ is now given by 
\begin{eqnarray}
a_{\sigma} = \chi u^{0}u^{0}k_{0}k_{0} - \chi u^{i}u^{i}k_{i}k_{i}+k_{\tau}^{2}+k_{x}^{2}+\left(
\frac{k_{x}a}{h}-\frac{2\pi n}{h}\right)  ^{2}+k_{z}^{2}+M_{\Phi}%
^{2},\label{ln}
\label{l4}%
\end{eqnarray}
where $M_{\Phi}^{2}=m^{2}+\frac{\lambda}{2}\Phi^{2}$ and $\sigma = (k_{\tau}, k_x, k_z, n)$ stands for the set of quantum numbers, with $ n=0,\pm1,\pm2,...$\;. In order to carry out further calculations, one has to specify the direction
of the constant 4-vector $u^{\mu}$. As we have explained before, the unit vector can be time-like, that is, $u^{t} = (1,0,0,0)$ or space-like, in which case we may have three possibilities: $u^{x}=(0,1,0,0)$, $u^{y}=(0,0,1,0)$ and $u^{x}=(0,0,0,1)$. Let us, next, analyze the time-like case first.

\subsection{Time-like case $u^{t}=(1,0,0,0)$}

Here we first consider the case where the 4-vector $u^{\mu}$ is a time-like
vector, i.e., $u^{t}=\left(  1,0,0,0\right)  $. Consequently, the eigenvalues
(\ref{ln}) are now written as,%
\begin{eqnarray}
a_{\sigma}  = \left(  1+\chi\right)  k_{\tau}^{2} +  k_{x}^{2}+\left(  \frac{k_{x}a}%
{h}-\frac{2\pi n}{h}\right)  ^{2}+k_{z}^{2}+M_{\Phi}^{2},\label{l5.1}
\end{eqnarray}
where it is clear that the Lorentz violation is in the time direction since the violation parameter $\chi$ is present only in the $\tau$-component part. The calculations required to construct the
Riemann zeta function from the eigenvalues (\ref{l5.1}) go in a similar way
as what was done in the previous section. Thus, we shall write only the main
results.

The resulting one-loop correction (\ref{z1}) for the case under consideration
is written as,%
\begin{equation}
V^{\left(  1\right)  }\left(  \Phi\right)  =\frac{M_{\Phi}^{4}}%
{64\pi^{2}\left(  1+\chi\right)^{\frac{1}{2}}}\left[  \ln\left(  \frac{M_{\Phi}%
^{2}}{\mu^{2}}\right)  -\frac{3}{2}\right]  -  \frac{M_{\Phi}^{4}}{2\pi^{2}\left(  1+\chi\right)^{\frac{1}{2}}}\sum_{j=1}^{\infty}f_{2}\left( jM_{\Phi
}\sqrt{a^2 + h^2}\right)  ,
\label{l8}%
\end{equation}
where $d$ is given in Eq.~(\ref{d1}) and the function $f_{\mu}(x)$ has been defined in
Eq.~(\ref{f}). Comparing the expression for the one-loop correction (\ref{l8})
with the previous result in Eq.~(\ref{v1m}), one can see that when the privileged
direction for the Lorentz violation to occur is the time direction, a modification
appears in the form of the term $\left(  1+\chi\right)^{\frac{1}{2}}$ in the
denominator.

\begin{figure}[h!]
	\centering
	{\includegraphics[width=8cm]{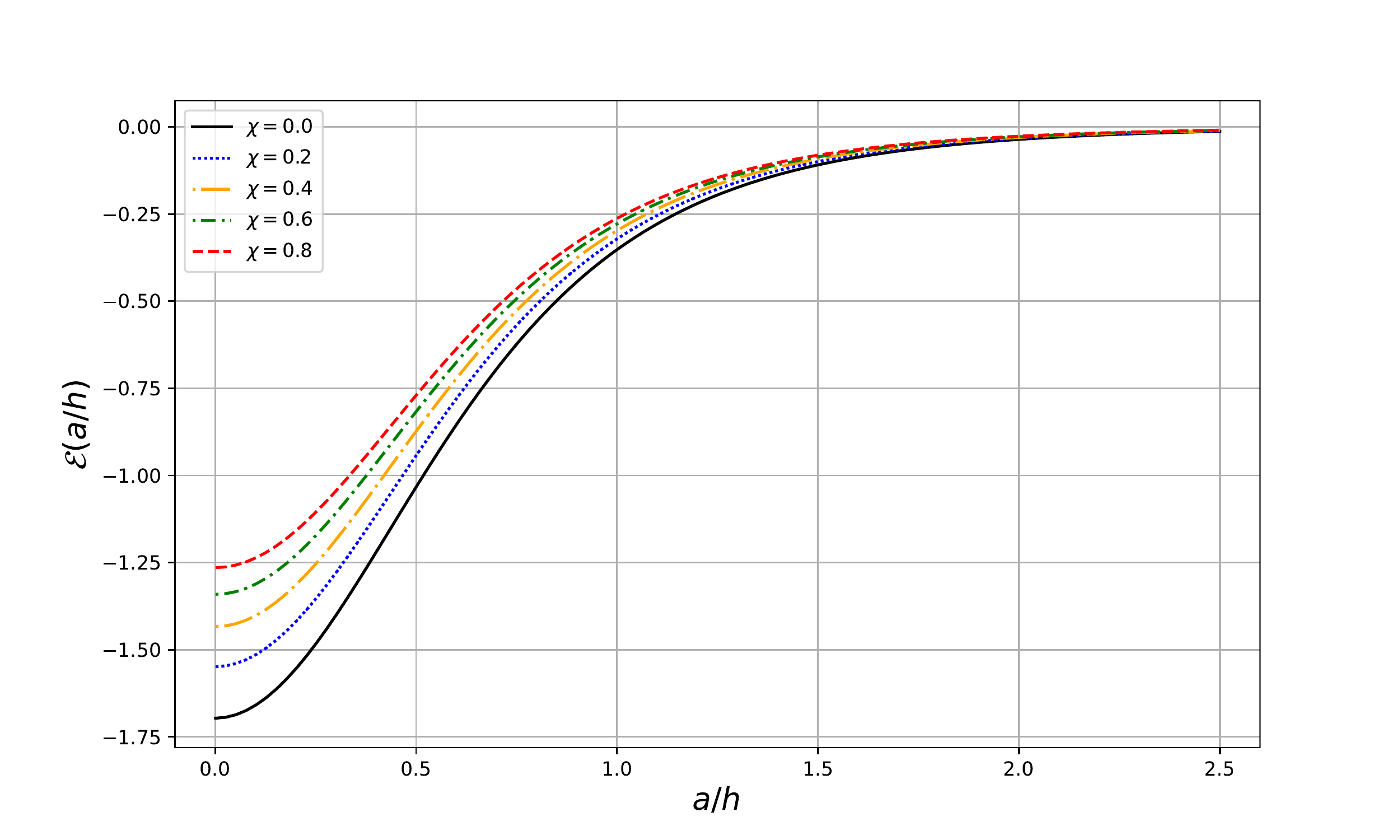} }%
	\caption{Plot of the dimensionless Casimir energy density $\mathcal{E} = \frac{2\pi^2}{m^4}\mathcal{E}_C$ with respect to $\frac{a}{h}$ from Eq. \eqref{l11}, for different values of $\chi$, considering the time-like Lorentz violation case. It is assumed $mh=1$ and $\lambda=10^{-2}$.}
	\label{figure2}%
\end{figure}

In order to find the renormalized effective potential, one have to use the
renormalization conditions (\ref{2.8}), (\ref{2.9}) and (\ref{2.11})\ in the Minkowski
limit $d\rightarrow\infty$ \cite{Toms:1979ij, Porfirio:2019gdy, Cruz:2020zkc}. These conditions lead to
the following renormalization constants:%
\begin{align}
\frac{C_{1}}{4!}  &  =\frac{\lambda^{2}}{256\pi^{2}\left(  1+\chi\right)^{\frac{1}{2}}}%
\ln\left(  \frac{\mu^{2}}{m^{2}}\right)  ,\nonumber\\
\frac{C_{2}}{2!}  &  =\frac{\lambda m^{2}}{64\pi^{2}\left(  1+\chi\right)^{\frac{1}{2}}}%
+\frac{\lambda m^{2}}{64\pi^{2}\left(  1+\chi\right)^{\frac{1}{2}}}\ln\left(
\frac{\mu^{2}}{m^{2}}\right)  ,\nonumber\\
C_{3}  &  =\frac{3m^{4}}{128\pi^{2}\left(  1+\chi\right)^{\frac{1}{2}}}+\frac
{m^{4}}{64\pi^{2}\left(  1+\chi\right)^{\frac{1}{2}}}\ln\left(  \frac{\mu^{2}%
}{m^{2}}\right)  . \label{l9}%
\end{align}
Knowing the form of the renormalization constants $C_{i}$, one can write the
renormalized effective potential up to one-loop as,%
\begin{align}
  V_{\text{eff}}^{R}\left(  \Phi\right) & =\frac{\lambda}{4!}\Phi^{4}+\frac{m^{2}}%
{2}\Phi^{2}+\frac{\lambda m^{2}\Phi^{2}}{64\pi^{2}\left(  1+\chi\right)^{\frac{1}{2}}}\left[  \ln\left(  \frac{M_{\Phi}^{2}}{m^{2}}\right)  -\frac{1}{2}\right]
 + \frac{\lambda^{2}\Phi^{4}}{256\pi^{2}\left(  1+\chi\right)^{\frac{1}{2}}%
}\left[  \ln\left(  \frac{M_{\Phi}^{2}}{m^{2}}\right)  -\frac{3}{2}\right]\nonumber\\
& +\frac{m^{4}}{64\pi^{2}\left(  1+\chi\right)^{\frac{1}{2}}}\ln\left(  \frac
{M_{\Phi}^{2}}{m^{2}}\right)  - \frac{M_{\Phi}^{4}}{2\pi^{2}\left(  1+\chi\right)^{\frac{1}{2}}}\sum_{j=1}^{\infty}f_{2}\left( jM_{\Phi
}\sqrt{a^2 + h^2}\right) . \label{l10}%
\end{align}
Furthermore, by taking the vacuum state, i.e., $\Phi=0$, in the above renormalized effective potential, we obtain the Casimir energy density which
is given by%
\begin{equation}
\mathcal{E}_{\text{C}}=\left.  V_{\text{eff}}^{R}\left(  \Phi\right)  \right\vert _{\Phi=0}= -  \frac{m^{4}}{2\pi^{2} \left(  1+\chi\right)^{\frac{1}{2}}}\sum_{j=1}^{\infty}f_{2}\left( jm\sqrt{a^2 + h^2}\right) . \label{l11}%
\end{equation}
Hence, the Casimir energy density changes by a factor of $\left(  1+\chi\right)
^{-\frac{1}{2}}$ when comparing Eq.~(\ref{l11}) to Eq. (\ref{casm}). 
In Fig.\ref{figure2}, Eq. \eqref{l11} is plotted as a dimensionless energy density $\frac{2\pi^2}{m^4}\mathcal{E}_{C}$ with respect to the dimensionless length $\frac{a}{h}$, for different values of the violation parameter $\chi$, also assuming $mh =1$ and $\lambda = 10^{-2}$. We can see that the effect of the Lorentz symmetry violation in the time direction is to increase the Casimir energy density. However, at large values of $\frac{a}{h}$, the Casimir energy density \eqref{l11} goes to zero, regardless the value of the violation parameter $\chi$. 
%
%

The Casimir energy density associated with the massless scalar field is obtained from Eq. \eqref{l11} in the limit of small arguments for the Macdonald function and is given by
\begin{equation}
\mathcal{E}_{\text{C}}=-\frac
{\pi^{2}}{90\left(  1+\chi\right)^{\frac{1}{2}}(h^2 + a^2)^{2}}. \label{cas0l}%
\end{equation}

\begin{figure}[h!]
	\centering
	{\includegraphics[width=8cm]{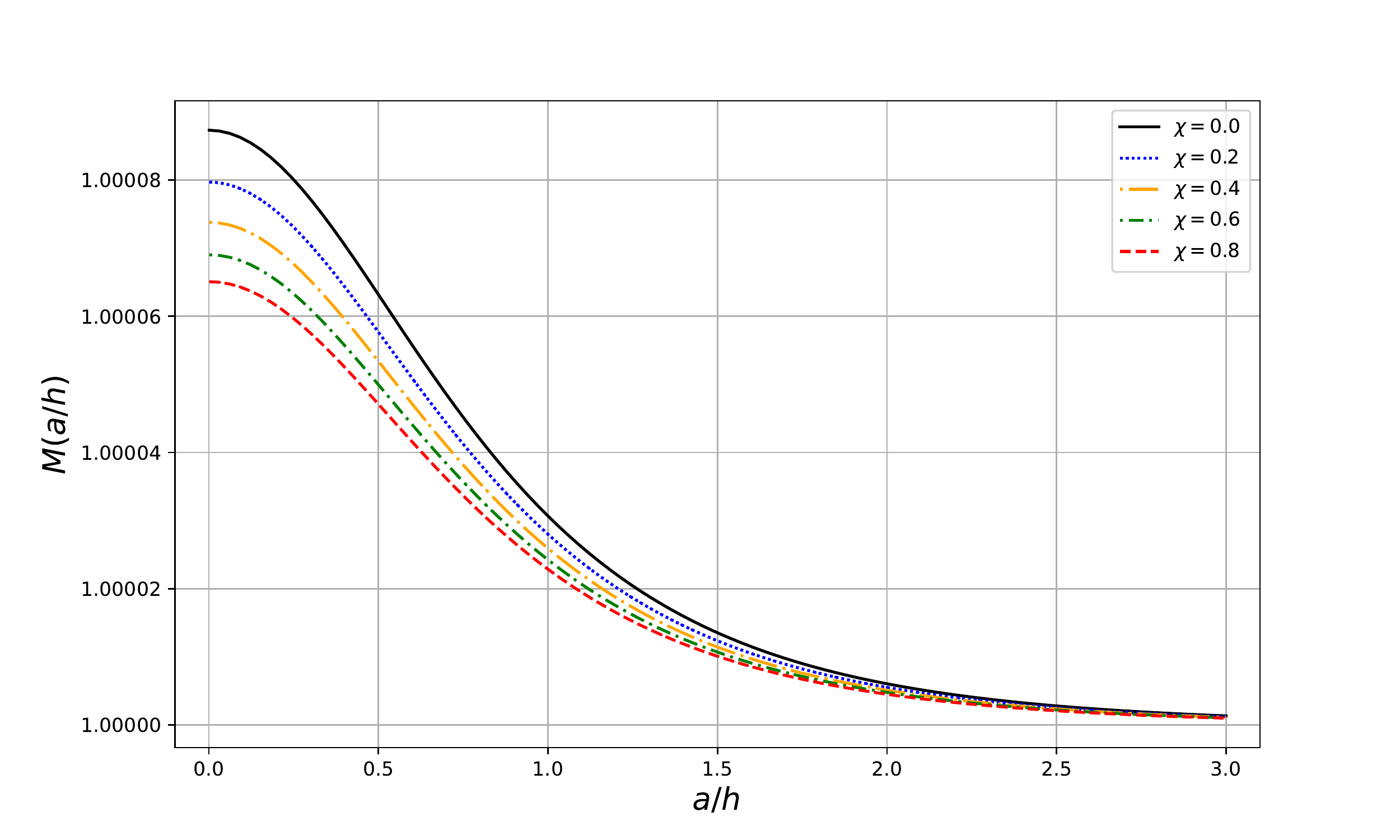} }%
	\qquad
	{\includegraphics[width=8cm]{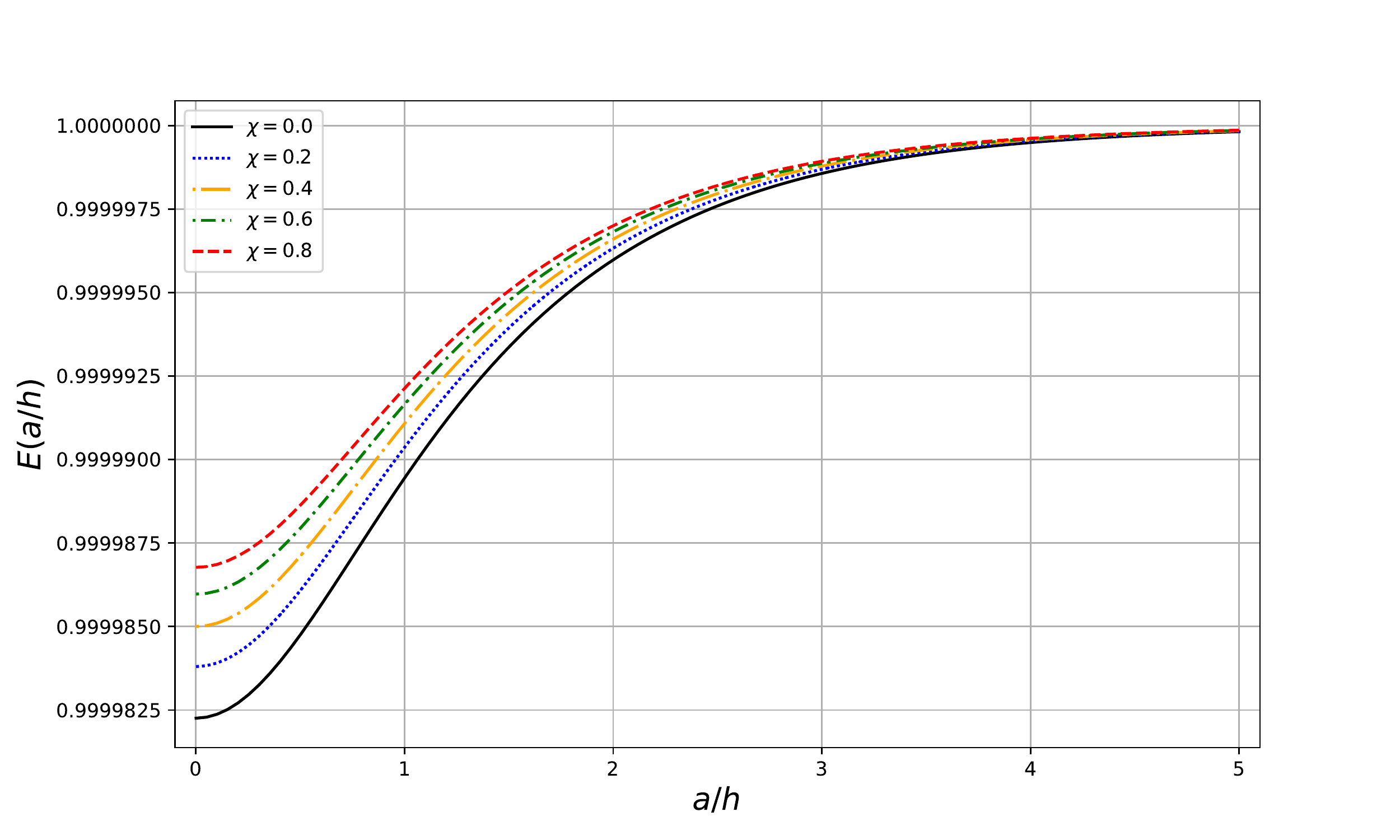} }%
	\caption{For different values of $\chi$, in the time-like Lorentz violation case, the plot on the left shows the curves of $M=\frac{m_{\text{T}}}{m}$ with respect to $\frac{a}{h}$ from Eq. \eqref{l12}, while the plot on the right shows the curves of $E=\frac{\mathcal{E}_{C}^{\lambda}}{\mathcal{E}_C}$ with respect to $\frac{a}{h}$ from Eq. \eqref{caslcviolation}. For both plots it is assumed $mh=1$ and $\lambda=10^{-2}$.}
	\label{figure3}%
\end{figure}

It is interesting to see how the Lorentz symmetry violation affects the
topological mass acquired by the field. Thus, using the condition (\ref{2.9})
for the renormalized effective potential (\ref{l10}), we obtain,%
\begin{equation}
m_{\text{T}}^{2}=m^{2}\left[  1+\frac{\lambda}{4\pi^{2}\left(  1+\chi\right)^{\frac{1}{2}}}\sum_{j=1}^{\infty}f_{1}\left( jm\sqrt{a^2 + h^2}\right)   \right]
. \label{l12}%
\end{equation}
As one can see, the correction for the mass is also multiplied by the factor
$\left(  1+\chi\right)  ^{-\frac{1}{2}}$. Hence, the Lorentz violation also affects the topological mass. On the left of Fig.\ref{figure3}, Eq. \eqref{l12} is plotted as a dimensionless mass $\frac{m_{\text{T}}}{m}$ with respect to the dimensionless length $\frac{a}{h}$, for different values of the violation parameter $\chi$, also assuming $mh =1$ and $\lambda=10^{-2}$. We can see that the effect of the Lorentz violation is to decrease the topological mass. However, the one-loop correction to the mass $m$ goes to zero at large values of $\frac{a}{h}$, making the mass $m$ dominant, regardless the value of $\chi$.

One can also take, from Eq. \eqref{l12}, the massless scalar field limit. Then, by considering the limit of small arguments of the Macdonald function, we have
\begin{equation}
m_{\text{T}}^{2}=\frac{\lambda}{24\left(  1+\chi\right)^{\frac{1}{2}}(a^2 + h^2)}, \label{mass0t}%
\end{equation}
which is a pure topological contribution, as a consequence of the one-loop correction.

Now we turn our attention to the two-loop correction. Taking the same steps
that leads to Eq.~(\ref{v2.4}), one can easily obtain the order-$\lambda$ loop correction
to the Casimir energy density \eqref{l11}. This contribution has the following form:%
\begin{equation}
\left.  V^{\left(  2\right)  }\left(  \Phi\right)  \right\vert _{\Phi
=0}=\frac{\lambda m^{4}}{32\pi^{4}(1 + \chi)}\left[  \sum_{j=1}^{\infty}%
f_1\left(  jm\sqrt{a^2 + h^2}\right)  \right]  ^{2}. \label{v2.4LV}%
\end{equation}
Therefore, the effect of the Lorentz violation to this correction
is codified in the factor $\left(  1+\chi\right)  ^{-1}$.

%

For the massless scalar field case, from Eq. \eqref{v2.4LV}, we obtain
\begin{equation}
\left.  V^{\left(  2\right)  }\left(  \Phi\right)  \right\vert _{\Phi
=0}=\frac{\lambda}{1152(a^2 + h^2)^2(1 + \chi)}.
\label{masslessviolation}%
\end{equation}

Finally, the Casimir energy density, with its loop correction of order $\lambda$, is given by
\begin{equation}
\mathcal{E}_{\text{C}}^{\lambda}= \mathcal{E}_C + \frac{\lambda m^{4}}{32\pi^{4}(1 + \chi)}\left[  \sum_{j=1}^{\infty}%
f_1\left(  jm\sqrt{a^2 + h^2}\right)  \right]  ^{2},
\label{caslcviolation}%
\end{equation}
for the massive scalar field case and 
\begin{equation}
\mathcal{E}_{\text{C}}^{\lambda}=  \mathcal{E}_C + \frac{\lambda}{1152(a^2 + h^2)^2(1 + \chi)}, 
\label{caslcmasslessviolation}%
\end{equation}
for the massless scalar field case. Note that the first term on the r.h.s of Eq. \eqref{caslcviolation} is given by \eqref{l11}, while the first term on the r.h.s of Eq. \eqref{caslcmasslessviolation} is given by \eqref{l12}. On the right of Fig.\ref{figure3}, Eq. \eqref{caslcviolation} is plotted as a dimensionless energy density $\frac{\mathcal{E}_{C}^{\lambda}}{\mathcal{E}_C}$ with respect to the dimensionless length $\frac{a}{h}$, for different values of the violation parameter $\chi$, also assuming $mh =1$ and $\lambda=10^{-2}$. We can see that the effect of the order-$\lambda$ loop correction is to increase the Casimir energy density \eqref{caslcviolation}. However, this correction goes to zero at large values of $\frac{a}{h}$, making the Casimir energy density \eqref{l11} the dominant one.
%
\subsection{Space-like case $u^{x}=\left(  0,1,0,0\right)  $}

Now we consider the case in which the privileged direction is the $x$ one.
Thus, the 4-vector $u^{\mu}$ is written as $u^{x}=\left(  0,1,0,0\right)  $.
Note that this affects directly the component which is related to the boundary
condition (\ref{he1}), in this case, the radius $a$ of the helix. Under this specification, the eigenvalues from
Eq.~(\ref{ln}) are written in the form,%
\begin{equation}
a_{\sigma}=k_{\tau}^{2}+\left(  1-\chi\right)  k_{x}^{2}+\left(  \frac{k_{x}%
a}{h}-\frac{2\pi n}{h}\right)  ^{2}+k_{z}^{2}+M_{\Phi}^{2}. \label{l15}%
\end{equation}
The calculation of the Riemann zeta function here follows similar steps as the ones in the previous sections. Thereby, considering the eigenvalues \eqref{l15}, the one-loop
correction to the effective potential is found to be%
\begin{equation}
V^{\left(  1\right)  }\left(  \Phi\right)  = \frac{M_{\Phi}^{4}}%
{64\pi^{2}\left(  1-\chi\right)^{\frac{1}{2}}}\left[  \ln\left(\frac{M_{\Phi}^{2}}%
{\mu^{2}}\right)-\frac{3}{2}\right]  - 
\frac{M_{\Phi
}^{4}}{2\pi^{2}\left(  1-\chi\right)^{\frac{1}{2}} }\sum_{j=1}^{\infty}f_{2}\left( jhM_{\Phi}\gamma_{x}\right)  , \label{l16}%
\end{equation}
where we have introduced the new notation
\begin{equation}
\gamma_{x}^{2}= 1 + r^2 \left(  1-\chi\right)^{-1}, \label{dx}%
\end{equation}
with $r=\frac{a}{h}$. Note that the one-loop correction \eqref{l16}, differently from the time-like case, is also affected by the violation parameter $\chi$ in the argument of the Macdonald function in the last term on the r.h.s. The sign of $\chi$ is also the opposite from the time-like case. Of course, these different aspects also appear in the expressions for the Casimir energy density, its loop correction and the topological mass, as we shall see below.

The renormalization constants $C_i$, as previously, are also affected by the violation parameter $\chi$. Again, they are found by using the renormalization conditions (\ref{2.8}), (\ref{2.9}) and
(\ref{2.11}). This gives 
\begin{align}
\frac{C_{1}}{4!}  &  = \frac{\lambda^{2}}{256\pi^{2}\left(  1-\chi\right)^{\frac{1}{2}}}%
\ln\left(\frac{\mu^{2}}{m^{2}}\right),\nonumber\\
\frac{C_{2}}{2!}  &  = \frac{\lambda m^{2}}{64\pi^{2}\left(  1-\chi\right)^{\frac{1}{2}}}%
+\frac{\lambda m^{2}}{64\pi^{2}\left(  1-\chi\right)^{\frac{1}{2}}}\ln\left(\frac{\mu
^{2}}{m^{2}}\right),\nonumber\\
C_{3}  &  =\frac{3m^{4}}{128\pi^{2}\left(  1-\chi\right)^{\frac{1}{2}}}+\frac
{m^{4}}{64\pi^{2}\left(  1-\chi\right)^{\frac{1}{2}}}\ln\left(\frac{\mu^{2}}{m^{2}}\right).
\label{l17}%
\end{align}

The important quantity here is the renormalized effective potential, up to one-loop order. In this sense, by considering the normalization constants in Eq. \eqref{l17} we find 
\begin{align}
V_{\text{eff}}^{R}\left(  \Phi\right)   &  = \frac{\lambda\Phi^{4}}{4!}+\frac
{m^{2}\Phi^{2}}{2}+\frac{m^{4}}{64\pi^{2}\left(  1-\chi\right)^{\frac{1}{2}} }%
\ln\left(\frac{M_{\Phi}^{2}}{m^{2}}\right)+\frac{\lambda m^{2}\Phi^{2}}{64\pi^{2}%
\left( 1-\chi\right)^{\frac{1}{2}} }\left[  \ln\left(\frac{M_{\Phi}^{2}}{m^{2}}\right)-\frac
{1}{2}\right] \nonumber\\
&  +\frac{\lambda^{2}\Phi^{4}}{256\pi^{2}\left(  1-\chi\right)^{\frac{1}{2}} %
}\left[  \ln\left(\frac{M_{\Phi}^{2}}{m^{2}}\right)-\frac{3}{2}\right]  - \frac{M_{\Phi
}^{4}}{2\pi^{2}\left(  1-\chi\right)^{\frac{1}{2}} }\sum_{j=1}^{\infty}f_{2}\left( jhM_{\Phi}\gamma_{x}\right)  . \label{l18}%
\end{align}
%
\begin{figure}[h!]
	\centering
	{\includegraphics[width=8cm]{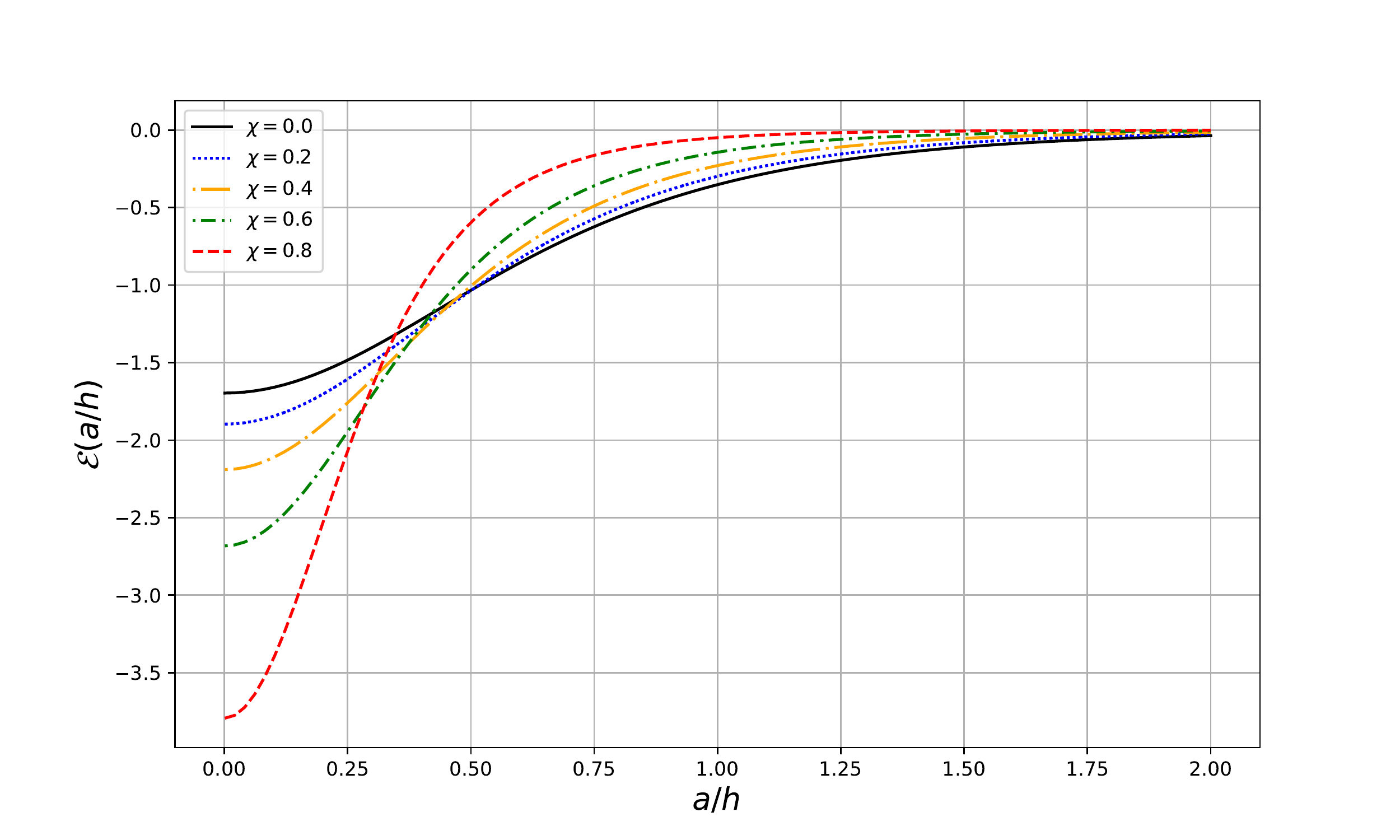} }%
	\caption{Plot of the dimensionless Casimir energy density $\mathcal{E} = \frac{2\pi^2}{m^4}\mathcal{E}_C$ with respect to $\frac{a}{h}$ from Eq. \eqref{caslx}, for different values of $\chi$, considering the Lorentz violation in the $x$-direction. It is assumed $mh=1$ and $\lambda=10^{-2}$.}
	\label{figure4}%
\end{figure}
%
Furthermore, by considering the vacuum state $\Phi=0$ in Eq. \eqref{l18}, the Casimir energy density when $u^{\mu}$ is taken in the
$x$-direction is given by
\begin{equation}
\mathcal{E}_C =\left.  V_{\text{eff}}^{R}\left(  \Phi\right)  \right\vert _{\Phi=0} = - \frac{m^{4}}{2\pi^{2}\left(  1-\chi\right)^{\frac{1}{2}} }\sum_{j=1}^{\infty}f_{2}\left( jmh\gamma_{x}\right)  . \label{caslx}%
\end{equation}
Comparing Eq.~(\ref{caslx}) with the Casimir energy density in the case when $u^{\mu
}$ is a time-like 4-vector in Eq.~(\ref{l11}), one can see the important role that
the direction of the Lorentz violation plays in the Casimir effect. In Fig.\ref{figure4}, Eq. \eqref{caslx} is plotted as a dimensionless energy density $\frac{2\pi^2}{m^4}\mathcal{E}_{C}$ with respect to the dimensionless length $\frac{a}{h}$, for different values of the violation parameter $\chi$, also assuming $mh =1$ and $\lambda = 10^{-2}$. We can see that the effect of the Lorentz symmetry violation in the $x$-direction is to initially decrease the Casimir energy density for some small values of $\frac{a}{h}$ and to increase for some other larger values. However, at even larger values of $\frac{a}{h}$, the Casimir energy density \eqref{caslx} goes to zero, regardless the value of the violation parameter $\chi$. 

The massless scalar field limit of the massive Casimir energy density \eqref{caslx} is written as
\begin{equation}
\mathcal{E}_C =-\frac
{\pi^{2}}{90h^4\gamma_{x}^{4}\left(  1-\chi\right)  ^{\frac{1}{2}}}, \label{caslx0}%
\end{equation}
which is affected by the Lorentz symmetry violation in a different way when compared with the time-like case in Eq. \eqref{cas0l}.

\begin{figure}[h!]
	\centering
	{\includegraphics[width=8cm]{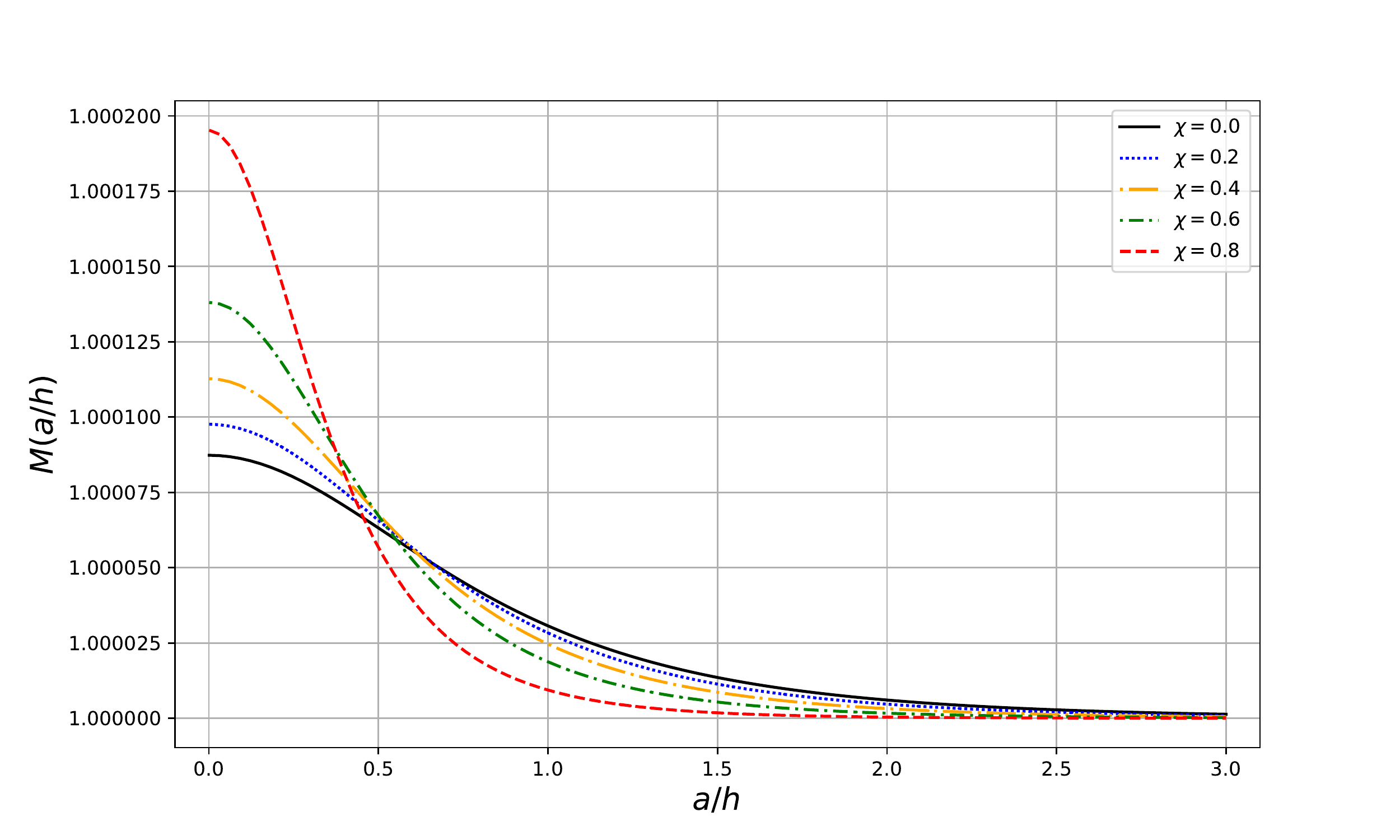} }%
	\qquad
	{\includegraphics[width=8cm]{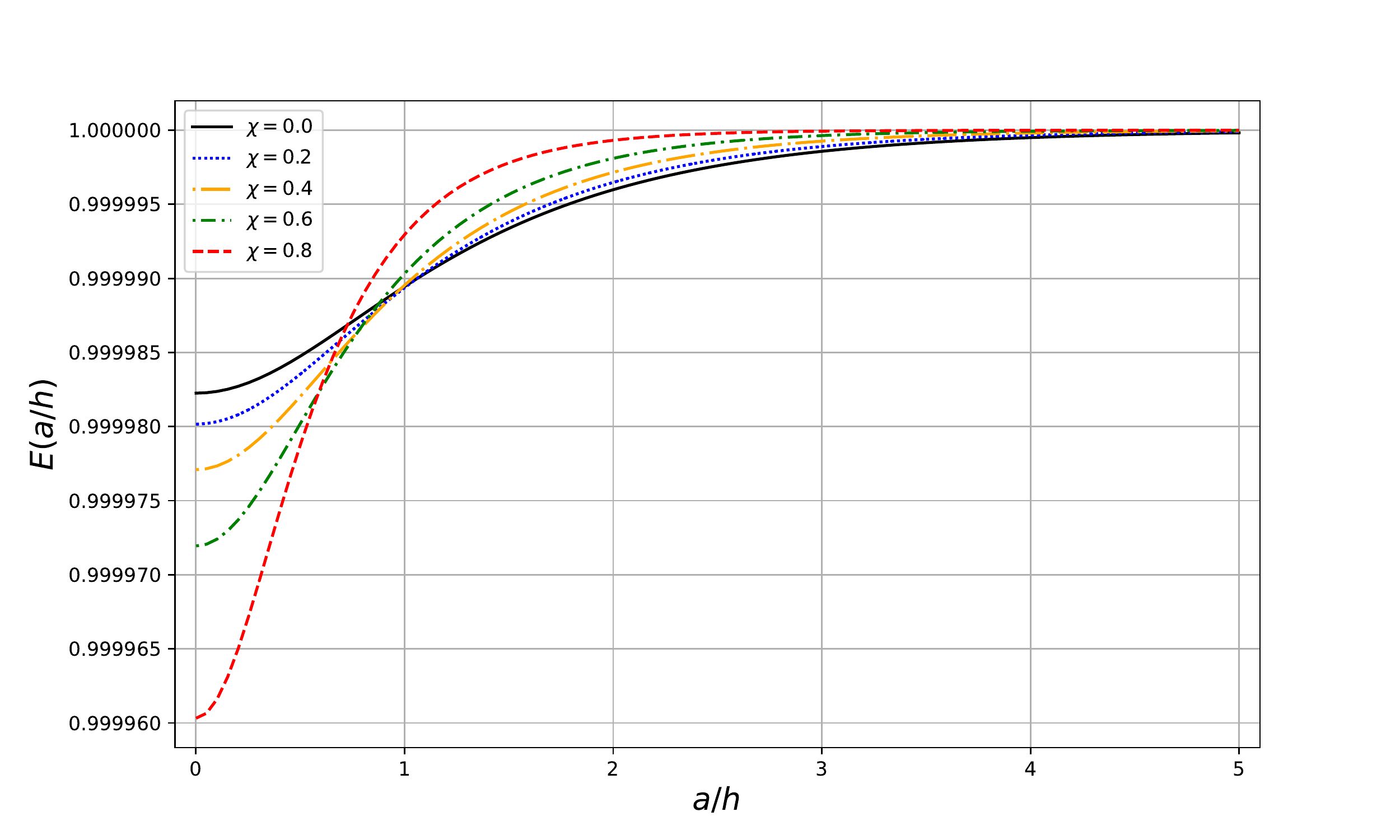} }%
	\caption{For different values of $\chi$, considering the Lorentz violation in the $x$-direction, the plot on the left shows the curves of $M=\frac{m_{\text{T}}}{m}$ with respect to $\frac{a}{h}$ from Eq. \eqref{l19}, while the plot on the right shows the curves of $E=\frac{\mathcal{E}_{C}^{\lambda}}{\mathcal{E}_C}$ with respect to $\frac{a}{h}$ from Eq. \eqref{spacelikex}. For both plots it is assumed $mh=1$ and $\lambda=10^{-2}$.}
	\label{figure5}%
\end{figure}

One can now consider the generation of
topological mass and how it is affected by the Lorentz symmetry violation in the $x$-direction. Hence, by using the condition (\ref{2.9}), along with the renormalized effective potential \eqref{l18}, one obtains the
following expression:%
\begin{equation}
m_{\text{T}}^{2}=m^{2}\left[  1+\frac{\lambda}{4\pi^{2}\left(  1-\chi\right)  ^{\frac{1}{2}} }\sum_{j=1}^{\infty
}f_{1}\left( jhm\gamma_{x}\right)
\right]  . \label{l19}%
\end{equation}
On the left of Fig.\ref{figure5}, Eq. \eqref{l19} is plotted as a dimensionless mass $\frac{m_{\text{T}}}{m}$ with respect to the dimensionless length $\frac{a}{h}$, for different values of the violation parameter $\chi$, also assuming $mh =1$ and $\lambda=10^{-2}$. We can see that the effect of the Lorentz violation is to initially increase the topological mass for some small values of $\frac{a}{h}$ and to decrease for some other larger values. However, at even larger values of $\frac{a}{h}$, the mass $m$ becomes dominant, regardless the value of the violation parameter $\chi$.
%
%

Moreover, by considering a massless scalar field, it acquires a pure topological mass in the
form,%
\begin{equation}
m_{\text{T}}  ^{2}=\frac{\lambda}
{24h^2\gamma_x^{2}(1 - \chi)^{\frac{1}{2}}}. \label{mass0x}%
\end{equation}
We note here that the last two results are completely different from the ones
obtained for the time-like case, given by
Eqs.~(\ref{l12}) and (\ref{mass0t}).

We can also consider the two-loop
contribution to the effective potential. By using the methods previously explained, and for the Lorentz violation case under consideration, the two-loop contribution at the vacuum state takes the form,%
\begin{equation}
V^{\left(  2\right)  }\left(  \Phi=0\right)  = \frac{\lambda m^{4}}%
{32\pi^{4}(1-\chi)}\left[  \sum_{j=1}^{\infty}f_1
\left(jhm\gamma_x\right)  \right]  ^{2}.
\label{l20}%
\end{equation}
As before, the massless scalar field limit follows from the above expression. It is written as
\begin{equation}
V^{\left(  2\right)  }\left(  \Phi=0\right)  = \frac{\lambda}%
{1152h^4\gamma_x^4(1-\chi)}.
\label{l211}%
\end{equation}

The complete expression for the Casimir energy density and its order-$\lambda$ loop correction is given by Eq. \eqref{caslx} plus Eq. \eqref{l20}, that is,
\begin{equation}
\mathcal{E}_{\text{C}}^{\lambda}=\mathcal{E}_C+ \frac{\lambda m^{4}}%
{32\pi^{4}(1-\chi)}\left[  \sum_{j=1}^{\infty}f_1
\left(jhm\gamma_x\right)  \right]  ^{2}.
\label{spacelikex}%
\end{equation}
Likewise, the complete expression in the massless scalar field case for the Casimir energy density and its order-$\lambda$ loop correction is given by Eq. \eqref{caslx0} plus Eq. \eqref{l211}, i.e.,
\begin{equation}
\mathcal{E}_{\text{C}}^{\lambda}=\mathcal{E}_C +  \frac{\lambda}%
{1152h^4\gamma_x^4(1-\chi)}.
\label{spacelikexmassless}%
\end{equation}
It is clear that the Lorentz symmetry violation in the $x$-direction affects the Casimir energy density and its loop correction in a completely different way and one of the main features is that the violation parameter also affects the argument of the Macdonald function present in the loop correction \eqref{l20}. On the right of Fig.\ref{figure5}, Eq. \eqref{spacelikex} is plotted as a dimensionless energy density $\frac{\mathcal{E}_{C}^{\lambda}}{\mathcal{E}_C}$ with respect to the dimensionless length $\frac{a}{h}$, for different values of the violation parameter $\chi$, also assuming $mh =1$ and $\lambda=10^{-2}$. We can see that the effect of the Lorentz symmetry violation in the $x$-direction is to initially decrease the Casimir energy density \eqref{spacelikex} for some small values of $\frac{a}{h}$ and to increase for some other larger values. However, at even larger values of $\frac{a}{h}$, the loop correction \eqref{l20} goes to zero, and the Casimir energy density \eqref{caslx} becomes dominant, regardless the value of the violation parameter $\chi$.

\subsection{Space-like case $u^{y}=\left(  0,0,1,0\right)  $}
%
Let us now turn our attention to the case where $u^{\mu}$ is in the $y$-direction, i.e., $u^{y}=\left(  0,0,1,0\right)  $. In this case the
eigenvalues (\ref{ln}) becomes,%
\begin{equation}
a_{\sigma}=k_{\tau}^{2}+k_{x}^{2}+\left(  1-\chi\right)  \left(  \frac{k_{x}%
a}{h}-\frac{2\pi n}{h}\right)  ^{2}+k_{z}^{2}+M_{\Phi}^{2}. \label{l21}%
\end{equation}
Essentially the Lorentz violation in the $y$-direction affects the pitch $h$ of the helix. The one-loop correction to the effective potential is obtained as in the previous sections and, in the present case, is given by Eq. \eqref{l16} replacing $\gamma_x$ by $\gamma_y$, which is defined in Eq. \eqref{l22}. The normalization constants $C_i$ are exactly the same as those in Eq. \eqref{l17}. 

The renormalized effective potential here obeys the same symmetry as the one-loop correction mentioned above. In other words, the renormalized effective potential when the Lorentz violation is in the $y$-direction is given by (\ref{l18}), replacing $\gamma_x$ by $\gamma_y$, i.e.,
\begin{align}
V_{\text{eff}}^{R}\left(  \Phi\right)   &  =\frac{\lambda\Phi^{4}}{4!}+\frac
{m^{2}\Phi^{2}}{2}+\frac{m^{4}}{64\pi^{2}\left(  1-\chi\right)^{\frac{1}{2}}}%
\ln\left(\frac{M_{\Phi}^{2}}{m^{2}}\right)+\frac{\lambda m^{2}\Phi^{2}}{64\pi^{2}%
\left(  1-\chi\right)^{\frac{1}{2}}}\left[  \ln\left(\frac{M_{\Phi}^{2}}{m^{2}}\right)-\frac
{1}{2}\right] \nonumber\\
&  +\frac{\lambda^{2}\Phi^{4}}{256\pi^{2}\left(  1-\chi\right)^{\frac{1}{2}}%
}\left[  \ln\left(\frac{M_{\Phi}^{2}}{m^{2}}\right)-\frac{3}{2}\right]  -  \frac
{M_{\Phi}^{4}}{2\pi^{2}(1-\chi)^{\frac{1}{2}}}\sum
_{j=1}^{\infty}f_{2}\left( jhM_{\Phi}\gamma_{y}%
\right)  . \label{vry}%
\end{align}
where
\begin{equation}
\gamma_{y}^{2}=\left(  1-\chi\right)^{-1} + r^{2}. \label{l22}%
\end{equation}

\begin{figure}[h!]
	\centering
	{\includegraphics[width=8cm]{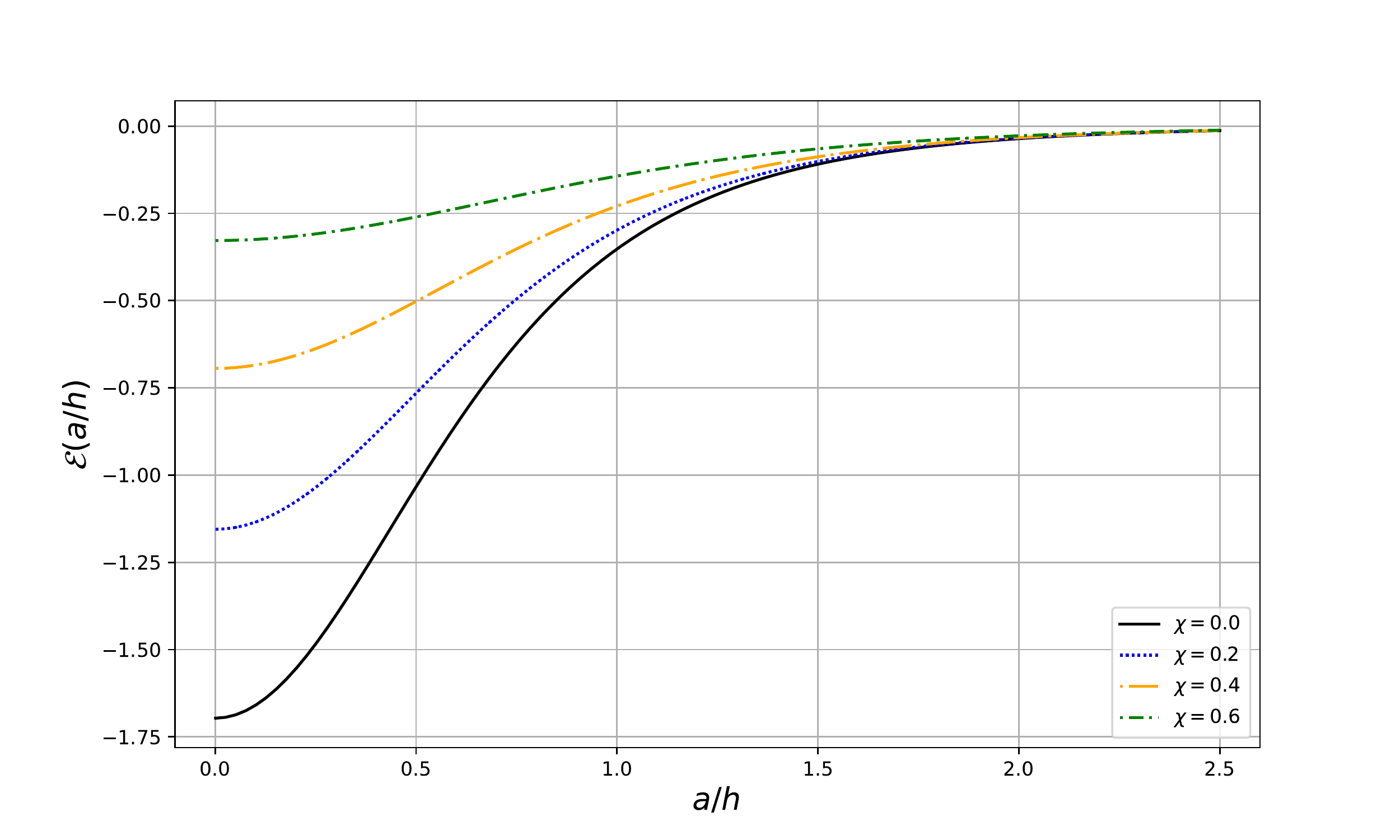} }%
	\caption{Plot of the dimensionless Casimir energy density $\mathcal{E} = \frac{2\pi^2}{m^4}\mathcal{E}_C$ with respect to $\frac{a}{h}$ from Eq. \eqref{casy}, for different values of $\chi$, considering the Lorentz violation in the $y$-direction. It is assumed $mh=1$ and $\lambda=10^{-2}$.}
	\label{figure6}%
\end{figure}

The Casimir energy density is, thereby, obtained by taking the vacuum state, $\Phi=0$, in the renormalized effective potential above, providing 
\begin{eqnarray}
\mathcal{E}_C = V_{\text{eff}}^{R}\left(  \Phi=0\right)  = -  \frac
{m^{4}}{2\pi^{2}(1-\chi)^{\frac{1}{2}}}\sum
_{j=1}^{\infty}f_{2}\left( jhm\gamma_{y}%
\right).
\label{casy}
\end{eqnarray}
In Fig.\ref{figure6}, Eq. \eqref{casy} is plotted as a dimensionless energy density $\frac{2\pi^2}{m^4}\mathcal{E}_{C}$ with respect to the dimensionless length $\frac{a}{h}$, for different values of the violation parameter $\chi$, also assuming $mh =1$ and $\lambda = 10^{-2}$. We can see that the effect of the Lorentz symmetry violation in the $y$-direction is to increase the Casimir energy density. However, for large values of $\frac{a}{h}$, the Casimir energy density \eqref{casy} goes to zero, regardless the value of the violation parameter $\chi$.

The massless scalar field limit of Eq. \eqref{casy} can be obtained as previously and, in this case, is given by
\begin{eqnarray}
\mathcal{E}_C =-\frac
{\pi^{2}}{90h^4\gamma_{y}^{4}\left(  1-\chi\right)  ^{\frac{1}{2}}}.
\label{cas0y}
\end{eqnarray}
Again, here we observe a symmetry when compared with the previous case, when the Lorentz violation occurs in the $x$-direction. That is, the Casimir energy density in the massless case is given by \eqref{caslx0} replacing $\gamma_x$ by $\gamma_y$. 

Of course the symmetry described above extends for the topological mass results (\ref{l19})
and (\ref{mass0x}). In the present case, these quantities are written as,%
\begin{eqnarray}
 m_{\text{T}}^{2} = m^{2}\left[  1+\frac{\lambda}{4\pi^{2}(1-\chi)^{\frac{1}{2}}}\sum_{j=1}^{\infty
}f_{1}\left( jmh\gamma_y\right)
\right],
\label{TMyd}
\end{eqnarray}
for the massive case and
\begin{figure}[h!]
	\centering
	{\includegraphics[width=8cm]{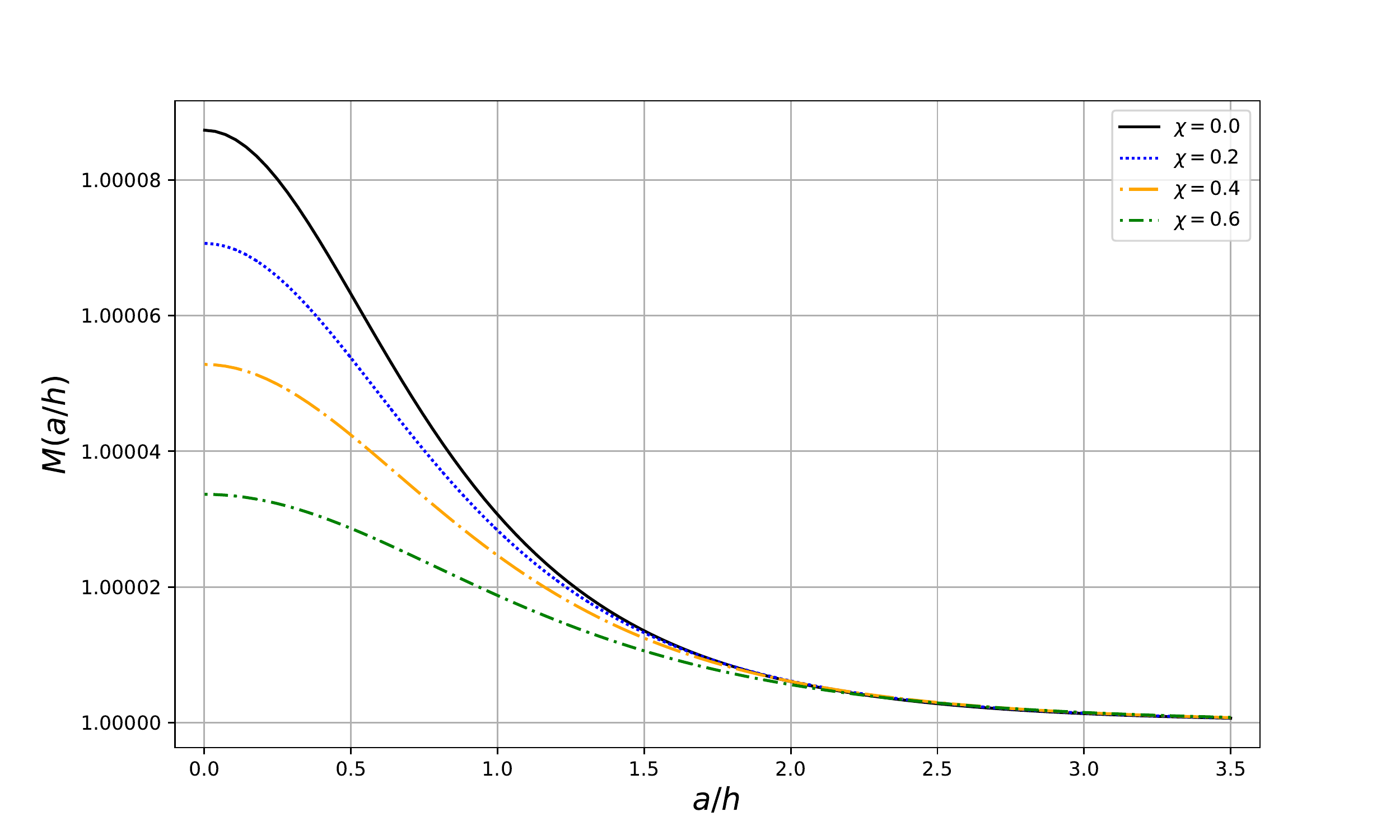} }%
	\qquad
	{\includegraphics[width=8cm]{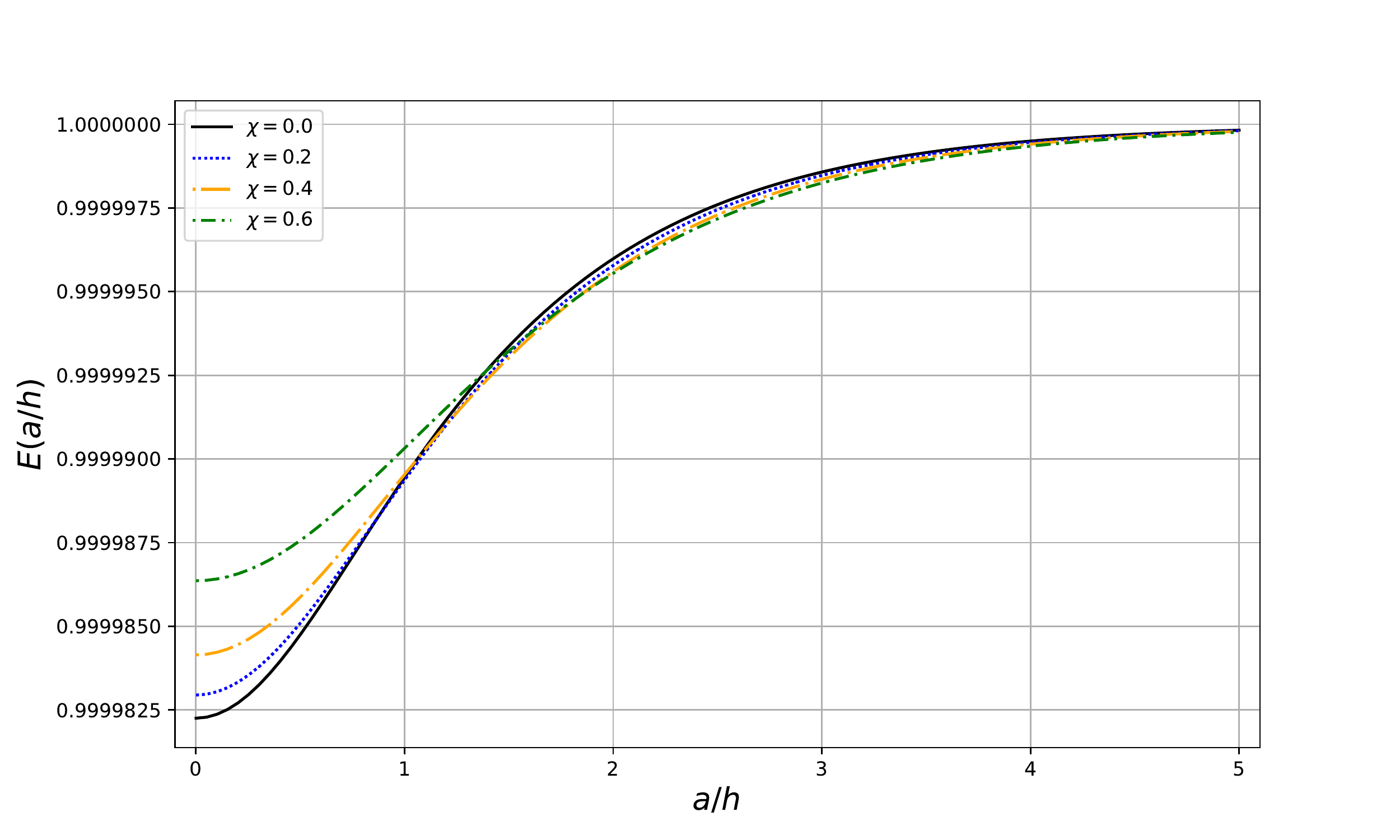} }%
	\caption{For different values of $\chi$, considering the Lorentz violation in the $y$-direction, the plot on the left shows the curves of $M=\frac{m_{\text{T}}}{m}$ with respect to $\frac{a}{h}$ from Eq. \eqref{TMyd}, while the plot on the right shows the curves of $E=\frac{\mathcal{E}_{C}^{\lambda}}{\mathcal{E}_C}$ with respect to $\frac{a}{h}$ from Eq. \eqref{spacelikexycase}. For both plots it is assumed $mh=1$ and $\lambda=10^{-2}$.}
	\label{figure7}%
\end{figure}
%
\begin{eqnarray}
 m_{\text{T}}^{2} = \frac{\lambda}{24h^2\gamma_y^2(1 - \chi)^{\frac{1}{2}}},
\end{eqnarray}
for the massless scalar field case. On the left of Fig.\ref{figure7}, Eq. \eqref{TMyd} is plotted as a dimensionless mass $\frac{m_{\text{T}}}{m}$ with respect to the dimensionless length $\frac{a}{h}$, for different values of the violation parameter $\chi$, also assuming $mh =1$ and $\lambda=10^{-2}$. We can see that the effect of the Lorentz violation is to decrease the topological mass. However, for large values of $\frac{a}{h}$, the mass $m$ becomes dominant, regardless the value of the violation parameter $\chi$.

Following the same steps as the ones in the previous sections, we find the order-$\lambda$ loop contribution to the Casimir energy density \eqref{casy} as%
\begin{equation}
V^{\left(  2\right)  }\left(  \Phi=0\right)  =\frac{\lambda m^{4}}%
{32\pi^{4}(1-\chi)}\left[  \sum_{j=1}^{\infty}f_1(jhm\gamma_y) \right]  ^{2},
\label{LCyd}
\end{equation}
for the massive case and
\begin{equation}
V^{\left(  2\right)  }\left(  \Phi=0\right)  = \frac{\lambda}%
{1152h^4\gamma_y^4(1-\chi)},
\label{masslessycase}%
\end{equation}
for the massless scalar field case. As expected, the existing symmetry between the Lorentz violation in the $x$ and $y$ directions is also present in the loop corrections \eqref{LCyd} and \eqref{masslessycase}.

The complete expression for the Casimir energy density \eqref{casy} plus its loop correction \eqref{LCyd} is written as
\begin{equation}
\mathcal{E}_{\text{C}}^{\lambda}= \mathcal{E}_C + \frac{\lambda m^{4}}%
{32\pi^{4}(1-\chi)}\left[  \sum_{j=1}^{\infty}f_1
\left(jhm\gamma_y\right)  \right]  ^{2},
\label{spacelikexycase}%
\end{equation}
for the massive case and 
\begin{equation}
\mathcal{E}_{\text{C}}^{\lambda}=\mathcal{E}_C +  \frac{\lambda}%
{1152h^4\gamma_y^4(1-\chi)}, 
\label{spacelikexmasslessycase}%
\end{equation}
for the massless scalar field case. On the right of Fig.\ref{figure7}, Eq. \eqref{spacelikexycase} is plotted as a dimensionless energy density $\frac{\mathcal{E}_{C}^{\lambda}}{\mathcal{E}_C}$ with respect to the dimensionless length $\frac{a}{h}$, for different values of the violation parameter $\chi$, also assuming $mh =1$ and $\lambda=10^{-2}$. We can see that the effect of the Lorentz symmetry violation in the $y$-direction is to initially increase the Casimir energy density \eqref{spacelikexycase} for some small values of $\frac{a}{h}$ and to slightly decrease for some other larger values. However, at even larger values of $\frac{a}{h}$, the loop correction \eqref{l20} goes to zero, and the Casimir energy density \eqref{casy} becomes dominant, regardless the value of the violation parameter $\chi$.

We end the discussion by stressing the fact that by considering a CPT-even aether-like Lorentz symmetry violation, all the relevant expressions, namely, the renormalized effective potential, the Casimir energy density, its order-$\lambda$ loop correction, and the generation of topological mass are affected by the violation parameter $\chi$. In particular, the cases where the Lorentz violation occurs in the $x$ and $y$ directions exhibit a symmetry. That is, we are able to obtain the expressions in one case from the expressions from the other case, by exchanging the parameters $\gamma_x$ or $\gamma_y$. Moreover, the case where the 4-vector $u^{\mu}$ points in the $z$-direction also exhibit a symmetry with respect to the Lorentz violation in the time direction. In other words, we can obtain the expressions for the renormalized effective potential, the Casimir energy density, its order-$\lambda$ loop correction, and the generation of topological mass from the time-like case by replacing $\chi$ by $-\chi$.

\section{Concluding remarks}
\label{sec5}
We have studied the Casimir effect arising from modifications of quantum vacuum fluctuations of a self-interacting scalar field subjected to a helix boundary condition in Minkowski spacetime preserving the Lorentz symmetry, as well as in a scenario with a CPT-even aether-type violation of Lorentz symmetry. In order to do that we have used the effective potential approach obtained via path integral method. We, thus, have calculated the renormalized effective potencial up to one-loop correction and show that when it is taken at the vacuum state, previous results for the Casimir energy density are recovered  \cite{QuantumSpringFromCasimir, QuantumSpring, QuantumSpringD+1, Aleixo:2021cfy}. At one-loop level, we have also shown that the mass of the self-interacting scalar field acquires a topological term of order $\lambda$, which is generated due to the helix boundary condition, something not obtained in literature before. We have, then, extended the results from Refs. \cite{QuantumSpringFromCasimir, QuantumSpring, QuantumSpringD+1, Aleixo:2021cfy} by analyzing how the Lorentz violation in each of the spacetime directions affects them and show the differences by plotting the expressions in Figs.\ref{figure2},\ref{figure4} and \ref{figure6}. The effect of the one-loop correction to the mass $m$ of the field is shown in Fig.\ref{figure1} (left side) preserving the Lorentz symmetry and in Figs.\ref{figure3},\ref{figure5} and \ref{figure7} (left side) by assuming the Lorentz symmetry is violated. 

In addition, we have also obtained the two-loop correction to the effective potential and show that, at the vacuum state, it is a correction of order $\lambda$ to the Casimir energy density considered in Refs. \cite{QuantumSpringFromCasimir, QuantumSpring, QuantumSpringD+1, Aleixo:2021cfy}. The effect of this loop correction on the Casimir energy density, compared to the one without correction, is plotted in Fig.\ref{figure1} (right side). Furthermore, we have analyzed the correction of order $\lambda$ to the Casimir energy density assuming the Lorentz violation in each one of the spacetime directions. The results are shown in the plots of Figs.\ref{figure3},\ref{figure5} and \ref{figure7} (right side). 

We would like to emphasize that the helicoidal geometry, considered here as a boundary condition, is present in a variate of structures in nature such as the DNA and cell membrane proteins. In the conclusions of Ref. \cite{Aleixo:2021cfy} a possible application has been pointed out. As the length dimensions of the helicoidal structure of the DNA or RNA is on the nanometers scale, the exactly scale where vacuum fluctuation effects become relevant, the Casimir energy density may induce genetic mutations on the DNA or RNA. Thus, adding up elements such as temperature corrections and also self-interaction, may offer a more realistic way to see how the genetic mutation is induced.

\appendix%

\section{Eigenvalues of the $\hat{A}$ operator}

\label{appendixa}

\renewcommand{\theequation}{A.\arabic{equation}} \setcounter{equation}{0}

In this appendix section we want to briefly analyze the four-dimensional volume, $\Omega _{4}$%
, and the eigenvalues of the operator $\hat{A}$ in Eq. (\ref{l}), when the helix
boundary condition is considered. Firstly, from Sec.\ref{sec3}, we note that the helix boundary condition leads to the following
eigenvalues constraint equation,%
\begin{equation}
k_{x}a-k_{y}h=2\pi n.  \label{ec}
\end{equation}%
Thus, the calculations presented in this paper have been performed by choosing to write the momentum $k_{y}$ in terms of
the momentum $k_{x}$, resulting in the eigenvalue equation in \eqref{aeigen}. As a consequence of this choice the four-dimensional volume, $%
\Omega _{4}$, taking into account the spacetime topology, is written as%
\begin{equation}
\Omega _{4}=\int d^{4}x=\Omega _{3}h,  \label{ec2}
\end{equation}%
where $\Omega _{3}$ is the three-dimensional volume associated with the
coordinates $\tau ,x,z$. However, we could equally well  have chosen to write the momentum $%
k_{x}$ in terms of the momemtum $k_{y}$. In this case, the eigenvalue equation in (\ref{aeigen2}) takes the following form,%
\begin{equation}
a_{\sigma }^{\prime }=k_{\tau }^{2}+k_{z}^{2}+k_{y}^{2}+\left( \frac{k_{y}h}{%
a}+\frac{2\pi n}{a}\right) ^{2}+M_{\Phi }^{2},\qquad \qquad n=0,\pm 1,\pm
2,....  \label{2.4}
\end{equation}%
For this choice, the four-dimensional volume is written as%
\begin{equation}
\Omega _{4}=\int d^{4}x=\Omega _{3}a,  \label{ec3}
\end{equation}%
where $\Omega _{3}$ is now the three-dimensional volume associated with the
coordinates $\tau ,y,z$. It is important to mention that the generalized
zeta function associated with the eigenvalues (\ref{2.4}) has the form
\begin{equation*}
\zeta _{a^{\prime }}\left( s\right) =\frac{\Omega _{3}}{\left( 2\pi \right)
^{3}}\int dk_{\tau }\ dk_{y}\ dk_{z}\sum_{n=-\infty }^{+\infty }\left[
k_{\tau }^{2}+k_{z}^{2}+k_{x}^{2}+\left( \frac{k_{y}h}{a}+\frac{2\pi n}{a}%
\right) ^{2}+M_{\Phi }^{2}\right] ^{-s},
\end{equation*}%
which, by adopting the same procedure as in Sec.\ref{sec3}, becomes
\begin{equation}
\zeta _{a^{\prime }}\left( s\right) =\frac{\Omega _{4}}{\left( 2\pi \right)
^{2}d}\frac{\pi ^{\frac{1}{2}}}{\Gamma \left( s\right) }\sum_{n=-\infty
}^{+\infty }\int_{0}^{\infty }d\tau \ \tau ^{2s-4}\exp \left\{ -\tau ^{2}%
\left[ M_{\Phi }^{2}+\frac{4\pi ^{2}}{d^{2}}n^{2}\right] \right\} .
\label{2.5}
\end{equation}%
Hence, the above zeta function is the same as the one presented in Eq.~(\ref{z}%
). We, thus, conclude that the results presented in this paper do not depend on the choice made in
how one writes the eigenvalue equation, if in terms of the momentum $k_x$ or in terms of the momentum $k_y$. Consequently, the results do not depend on whether we choose $\Omega_4=\Omega_3 h$ or $\Omega_4=\Omega_3 a$ for the four-dimensional Euclidian volume describing our model.

\section{Relation with the periodic boundary condition case}

\label{appendixb}

\renewcommand{\theequation}{B.\arabic{equation}} \setcounter{equation}{0}

In this section we show that the generalized zeta function, exhibited in
Eq.~(\ref{gzeta}), can be obtained from the generalized zeta function for the case
of a periodic boundary condition. This can also be done
for the case where Lorentz violation is considered, as we shall see.

Let us now remember that the generalized zeta function, considering helix boundary condition in a
scenario with no Lorentz violation is written as%
\begin{equation}
\zeta \left( s\right) =\frac{\Omega _{3}}{\left( 2\pi \right) ^{3}}\int
dk_{\tau }\ dk_{x}\ dk_{z}\sum_{n=-\infty }^{+\infty }\left[ k_{\tau
}^{2}+k_{x}^{2}+k_{z}^{2}+\left( \frac{k_{x}a}{h}-\frac{2\pi n}{h}\right)
^{2}+M_{\Phi }^{2}\right] ^{-s}.  \label{zhelixnl}
\end{equation}%
By performing the following change of variable in Eq.~(\ref%
{zhelixnl}), that is,%
\begin{equation}
k_{x}=\frac{h}{d}\left( p_{x}+\frac{2a\pi n}{hd}\right) ,\ \quad\qquad\qquad
d^{2}=h^{2}+a^{2},  \label{t1}
\end{equation}%
one obtains the zeta function as follows,%
\begin{equation}
\zeta \left( s\right) =\frac{h}{d}\frac{\Omega _{3}}{\left( 2\pi \right) ^{3}%
}\int dk_{\tau }\ dp_{x}\ dk_{z}\sum_{n=-\infty }^{+\infty }\left[ k_{\tau
}^{2}+p_{x}^{2}+k_{z}^{2}+\left( \frac{2\pi n}{d}\right) ^{2}+M_{\Phi }^{2}%
\right] ^{-s}.  \label{2.3}
\end{equation}%
The above result is to be associated with the generalized zeta function for
the case where a scalar field obeys a periodic boundary condition, with period $d$. The latter, in turn, is defined as
\begin{equation}
\zeta _{\mathrm{periodic}}\left( s,d\right) =\frac{\Omega _{3}}{\left( 2\pi
\right) ^{3}}\int dk_{\tau }\ dp_{x}\ dk_{z}\sum_{n=-\infty }^{+\infty }%
\left[ k_{\tau }^{2}+p_{x}^{2}+k_{z}^{2}+\left( \frac{2\pi n}{d}\right)
^{2}+M_{\Phi }^{2}\right] ^{-s}.  \label{zperiodic}
\end{equation}%
As a consequence, Eqs. (\ref{2.3}) and (\ref{zperiodic}) are connected by the
following relation,%
\begin{equation}
\zeta \left( s\right) =\frac{h}{d}\zeta _{\mathrm{periodic}}\left(
s,d\right) .  \label{ft1}
\end{equation}
In Ref. \cite{Porfirio:2019gdy}, the generalized zeta function associated with a periodic boundary condition has been obtained and is given by
\begin{equation}
\zeta _{\mathrm{periodic}}\left( s,d\right) =d\frac{\Omega _{3}M_{\Phi
}^{4-2s}}{16\pi ^{2}}\frac{\Gamma \left( s-2\right) }{\Gamma \left( s\right) 
}+d\frac{\Omega _{3}}{2^{s}\pi ^{2}\Gamma \left( s\right) }\left( \frac{%
M_{\Phi }}{d}\right) ^{2-s}\sum_{j=1}^{\infty }j^{s-2}K_{\left( 2-s\right)
}\left( jM_{\Phi }d\right) .  \label{zporf}
\end{equation}
Moreover, for the helix boundary condition, let us also remember the final form of the
generalized zeta function, i.e.,%
\begin{equation}
\zeta (s)=h\frac{\Omega _{3}M_{\Phi }^{4-2s}}{16\pi ^{2}}\frac{\Gamma \left(
s-2\right) }{\Gamma \left( s\right) }+h\frac{\Omega _{3}}{2^{s}\pi
^{2}\Gamma (s)}\left( \frac{M_{\Phi }}{d}\right) ^{2-s}\sum_{j=1}^{\infty
}j^{s-2}K_{\left( 2-s\right) }\left( jM_{\Phi }d\right) .  \label{zf2}
\end{equation}%
Therefore, we can see that by substituting the zeta function \eqref{zporf} in Eq. \eqref{ft1} we obtain the zeta function in Eq. \eqref{zf2}, for the helix boundary condition. Thus, we can reduce the problem analyzed here to the problem of a scalar field obeying a periodic boundary condition, with period $d$.

We can show a similar relation also in the case a Lorentz symmetry violation is considered. In this sense, we begin by considering that the Lorentz violation occurs in the $%
\tau $-direction. In this case, the generalized zeta function, after performing the transformation in Eq. \eqref{t1}, along with $\sqrt{\left( 1+\chi \right) }k_{\tau }=p_{\tau }$, we find 
\begin{equation}
\zeta _{L\tau }\left( s\right) =\frac{h}{d}\frac{\Omega _{3}}{\left( 2\pi
\right) ^{3}\sqrt{\left( 1+\chi \right) }}\int dp_{\tau }\ dp_{x}\ dk_{z}\sum_{n=-\infty }^{+\infty }%
\left[p_{\tau }^{2}+k_{z}^{2}+p_{x}^{2}+\left( \frac{%
2a\pi n}{d}\right) ^{2}+M_{\Phi }^{2}\right] ^{-s}.  \label{zlt}
\end{equation}%
Thereby, we can write the relation 
\begin{equation}
\zeta _{L\tau }\left( s\right) =\frac{h}{\sqrt{\left( 1+\chi \right) }d}%
\zeta _{\mathrm{periodic}}\left( s,d\right),
\label{tauD}
\end{equation}%
where $\zeta _{\mathrm{periodic}}\left( s,d\right)$ is given by Eq. \eqref{zporf}. As we can see, the zeta function for the case where the Lorentz violation occurs in the $%
\tau $-direction can be obtained from the zeta function for a periodic condition, with period $d$.  Note that, by symmetry, the relation considering the case of a Lorentz violation in the $z$-direction
is the same as in the case presented in Eq. \eqref{tauD}, only by making the change $\chi\rightarrow -\chi$.

Furthermore, by considering that the Lorentz violation is in the $x$-direction, the zeta function is
now written as%
\begin{equation}
\zeta _{Lx}\left( s\right) =\frac{\Omega _{3}}{\left( 2\pi \right) ^{3}}\int
dk_{\tau }\ dk_{z}\ dk_{x}\sum_{n=-\infty }^{+\infty }\left[ k_{\tau
}^{2}+k_{z}^{2}+\left( 1-\chi \right) k_{x}^{2}+\left( \frac{k_{x}a}{h}-%
\frac{2\pi n}{h}\right) ^{2}+M_{\Phi }^{2}\right] ^{-s}.  \label{1}
\end{equation}%
We can apply the following changing of variables in the above zeta function, that is,%
\begin{equation}
\xi k_{x}=\frac{\xi hp_{x}}{\sqrt{\xi ^{2}h^{2}+a^{2}}}+\frac{2\pi na\xi }{%
\xi ^{2}h^{2}+a^{2}},\ \qquad\qquad\qquad \xi =\sqrt{\left( 1-\chi \right) }.  \label{2}
\end{equation}%
This results in the relation%
\begin{equation}
\zeta _{Lx}\left( s\right) =\frac{h}{\xi d_{x}}\zeta _{\mathrm{periodic}%
}\left( s,d_{x}\right) ,\ \qquad\qquad\qquad d_{x}=\sqrt{h^{2}+\frac{a^{2}}{\xi ^{2}}}.
\label{2.6}
\end{equation}
Hence, this shows that the generalized zeta function \eqref{1}, for the heliz boundary condition, can be obtained from the generalized zeta function for a periodic condition, with period $d_x$.

Finally, let us consider the zeta function for the case in which a Lorentz violation occurs 
in the $y$-direction. The generalized zeta function in this case is written in the form,%
\begin{equation}
\zeta _{Ly}\left( s\right) =\frac{\Omega _{3}}{\left( 2\pi \right) ^{3}}\int
dk_{\tau }\ dk_{z}\ dk_{x}\sum_{n=-\infty }^{+\infty }\left[ k_{\tau
}^{2}+k_{x}^{2}+\left( 1-\chi \right) \left( \frac{k_{x}a}{h}-\frac{2\pi n}{h%
}\right) ^{2}+k_{z}^{2}+M_{\Phi }^{2}\right] ^{-s}.
\label{yD}
\end{equation}%
Now consider the transformation
\begin{equation}
k_{x}=\frac{hp_{x}}{\sqrt{h^{2}+a^{2}\xi ^{2}}}+\frac{2\pi na\xi ^{2}}{%
h^{2}+\xi ^{2}a^{2}}.
\end{equation}%
which by applying it in Eq. \eqref{yD} provides
\begin{equation}
\zeta _{Ly}\left( s\right) =\frac{h}{\xi d_{y}}\zeta _{\mathrm{periodic}%
}\left( s,d_{y}\right) ,\ \qquad\qquad\qquad d_{y}=\sqrt{a^{2}+\frac{h^{2}}{\xi ^{2}}.}
\label{2.7}
\end{equation}
Again, the zeta function \eqref{yD} for the helix boundary condition can be obtained from the zeta function for a periodic condition, with period $d_y$.

Therefore, we conclude this appendix section by saying that an alternative way for
obtaining the zeta function for the case of a helix boundary condition, is to
perform the change of variables presented above and show
that the results are given in terms of a generalized zeta function for a periodic condition.
%
{\acknowledgments}
A.J.D.F.J would like to thank the Brazilian agency CAPES for financial support. The author H.F.S.M. is partially supported by the Brazilian agency CNPq under Grant No. 311031/2020-0.

\end{document}